\documentclass[aps,preprint,amsfonts,amssymb,amsmath,eqsecnum,%
tightenlines,nofootinbib,showpacs,showkeys]{revtex4}

\newcommand{\cA}{\mathcal{A}}
\newcommand{\cC}{\mathcal{C}}
\newcommand{\cJ}{\mathcal{J}}
\newcommand{\cO}{\mathcal{O}}
\newcommand{\cP}{\mathcal{P}}
\newcommand{\cQ}{\mathcal{Q}}
\newcommand{\cT}{\mathcal{T}}

\newcommand{\fD}{\mathfrak{D}}
\newcommand{\fE}{\mathfrak{E}}
\newcommand{\fF}{\mathfrak{F}}
\newcommand{\fH}{\mathfrak{H}}
\newcommand{\fL}{\mathfrak{L}}
\newcommand{\fM}{\mathfrak{M}}
\newcommand{\fN}{\mathfrak{N}}
\newcommand{\fR}{\mathfrak{R}}
\newcommand{\fS}{\mathfrak{S}}
\newcommand{\bH}{\mathbf{H}}
\newcommand{\bK}{\mathbf{K}}
\newcommand{\bbZ}{\mathbb{Z}}
\newcommand{\bbR}{\mathbb{R}}
\newcommand{\bbC}{\mathbb{C}}
\newcommand{\rmi}{\mathrm{i}}
\newcommand{\rme}{\mathrm{e}}
\newcommand{\rmd}{\mathrm{d}}
\newcommand{\del}{\partial}
\newcommand{\braket}[1]{\bigl\langle{#1}\bigr\rangle}
\newcommand{\limaint}{\lim_{a\to\infty}\int_{-a}^{a}}
\newcommand{\Gaminf}{\Gamma_{\infty}}
\newtheorem{Def}{Definition}
\newtheorem{Pro}[Def]{Proposition}
\newtheorem{The}[Def]{Theorem}

\newtheorem{Cor}[Def]{Corollary}
\newcommand{\Ker}{\operatorname{Ker}}
\newcommand{\im}{\operatorname{Im}}

\begin{document}



%
%

\title{General Aspects of $\cP\cT$-Symmetric and
 $\cP$-Self-Adjoint Quantum Theory in a {K}rein Space}
\author{Toshiaki Tanaka}
\email{ttanaka@mail.ncku.edu.tw}
\affiliation{Department of Physics, Tamkang University,\\
 Tamsui 25137, Taiwan, R.O.C.}
\altaffiliation{Present address: Department of Physics,
 National Cheng-Kung University, Tainan 701, Taiwan, R.O.C.;
 National Center for Theoretical Sciences, Taiwan, R.O.C.}


\begin{abstract}

In our previous work, we proposed a mathematical framework for
$\cP\cT$-symmetric quantum theory, and in particular constructed
a Krein space in which $\cP\cT$-symmetric operators would naturally
act. In this work, we explore and discuss various general
consequences and aspects of the theory defined in the Krein space,
not only spectral property and $\cP\cT$ symmetry breaking but also
several issues, crucial for the theory to be physically acceptable,
such as time evolution of state vectors, probability interpretation,
uncertainty relation, classical-quantum correspondence, completeness,
existence of a basis, and so on. In particular, we show that
for a given real classical system we can always construct the
corresponding $\cP\cT$-symmetric quantum system, which indicates
that $\cP\cT$-symmetric theory in the Krein space is another
quantization scheme rather than a generalization of the traditional
Hermitian one in the Hilbert space. We propose a postulate for an
operator to be a physical observable in the framework.

\end{abstract}






\maketitle

\section{Introduction}
\label{sec:intro}

More than half a century ago, Dyson conjectured that the perturbation
series in the coupling constant $e^{2}$ in quantum electrodynamics
would be divergent by the physical argument that the theory with
$e^{2}<0$ where like charges attract is unstable against
the spontaneous pair creation of $e^{+}e^{-}$ and thus cannot have
a stable vacuum in contrast to the ordinary theory with $e^{2}>0$
\cite{Dy52}. It has been well-known now that in most of quantum
systems perturbation series are indeed divergent and at most
asymptotic, see e.g. Ref.~\cite{GZ90}. On the other hand, it might
not have been duly recognized that the Dyson's reasoning itself,
which led to the divergence of the perturbation series, is in general
invalid.

Immediately after the pioneering semi-classical analysis by Bender and
Wu \cite{BW68,BW69}, Simon in 1970 clarified with the mathematically
rigorous treatment the analytic structure of the energy eigenvalue
$E(g)$ of the quantum mechanical quartic anharmonic oscillator
\cite{Si70}
\begin{align}
\left(-\del^{2}+x^{2}+gx^{4}\right)\psi(x;g)=E(g)\psi(x;g).
\label{eq:AHO}
\end{align}
Besides the fact that the point $g=0$ is indeed a singularity of
$E(g)$, it was proved that $E(g)$ is analytic in the whole cut plane
$|\arg g|<\pi$ and in particular the theory (\ref{eq:AHO}) is
well-defined also for $g<0$. In other words, the system (\ref{eq:AHO})
with $g<0$ is \emph{mathematically stable} as an eigenvalue problem
although it is \emph{physically unstable} in the sense that the energy
eigenvalue $E(g)$ has a non-zero imaginary part besides its apparent
unstable shape of the potential. The underlying crucial fact is that
the analytic continuation of the system (\ref{eq:AHO}) with $g>0$
into the complex $g$ plane inevitably accompanies the rotation of
the domain $\bbR$ into the complex $x$ plane on which the theory
is defined. This is because the eigenfunctions $\psi(x;g)$ for
$g>0$ are normalizable in the sector
$|\arg(\pm x)+\frac{1}{6}\arg g|<\frac{\pi}{6}$ when
$|x|\to\infty$. As a consequence, the theory (\ref{eq:AHO}) with
$g<0$ obtained by the analytic continuation from $g>0$ is defined,
e.g., in the sectors $-\frac{4\pi}{3}<\arg x<-\pi$ and
$-\frac{\pi}{3}<\arg x<0$ ($|x|\to\infty$) when $\arg g=\pi$,
in the sectors $-\pi<\arg x<-\frac{2\pi}{3}$ and $0<\arg x<
\frac{\pi}{3}$ ($|x|\to\infty$) when $\arg g=-\pi$, and so on.
Hence, the important lesson drawn from the above fact is that we
must always take into account as well the effect on the linear space
on which a theory is defined when we change the sign of a parameter
involved in the Hamiltonian or Lagrangian.

This lesson was however not duly exercised when in 1973
Symanzik proposed $\lambda\phi^{4}$-theory with $\lambda<0$, which
is a quantum field theoretical version of the system (\ref{eq:AHO})
with $g<0$ in $1+3$ space-time dimension, as the first example of
an asymptotically free theory \cite{Sy73}; the majority considered
it physically unacceptable based on the intuition that it must be
unstable, though the investigation into this controversial model
has still persisted, e.g., Refs.~\cite{IIM75,Br76,GK85,St85,LR98}
(for a historical survey from a new viewpoint, see Ref.~\cite{Kl06}).

Recently, it was revealed that the model (\ref{eq:AHO}) with $g<0$
admits another novel treatment totally different from the analytic
continuation from the sector $g>0$. In 1998, Bender and Boettcher,
motivated by the Bessis--Zinn-Justin conjecture that the spectrum
of the Hamiltonian $H=p^{2}+x^{2}+\rmi x^{3}$ is real and positive,
found numerically that a family of the system
\begin{align}
H=p^{2}+m^{2}x^{2}+x^{2}(\rmi x)^{\epsilon}
\label{eq:PTAHO}
\end{align}
has indeed a real and positive spectrum for all $\epsilon\geq 0$,
and argued that it is the underlying $\cP\cT$ symmetry that ensures
these spectral properties \cite{BB98a,BBM99}. Here we note
especially the fact that although the latter model, when
$\epsilon=2$, looks the same as the system (\ref{eq:AHO}) with $g<0$,
their spectral properties are different. The key ingredient
underlying the difference is the different boundary conditions.
The regions where the eigenfunctions of the latter $\cP\cT$-symmetric
model (\ref{eq:PTAHO}) are normalizable, when it is defined for
$\epsilon>0$ as the continuation from the harmonic oscillator at
$\epsilon=0$, are given by the following sectors ($|x|\to\infty$)
\cite{BB98a}:
\begin{align}
\arg x=-\pi+\frac{\epsilon\pi}{2(4+\epsilon)}\pm\frac{\pi}{4+\epsilon}
 \quad\text{and}\quad\arg x=-\frac{\epsilon\pi}{2(4+\epsilon)}
 \pm\frac{\pi}{4+\epsilon}.
\end{align}
In particular, they are given by $-\pi<\arg x<-\frac{2\pi}{3}$ and
$-\frac{\pi}{3}<\arg x<0$ ($|x|\to\infty$) in the case $\epsilon=2$,
and are different from those for the quartic oscillator (\ref{eq:AHO})
with $\arg g=\pm\pi$. Hence, the $\cP\cT$-symmetric system
(\ref{eq:PTAHO}) with $\epsilon=2$ cannot be obtained by the analytic
continuation of the system (\ref{eq:AHO}) with $g>0$, a situation
anticipated by Symanzik~\cite{Sy73}.

In addition to the novel spectral properties, it was revealed that
the $\cP\cT$-symmetric model (\ref{eq:PTAHO}) admits a non-zero
vacuum expectation value $\langle 0|x|0\rangle$ even when $m^{2}>0$
and $\epsilon=2$ \cite{BM97,BMS00,BMY01}. All the non-perturbative
calculations in these papers indicate that the vacuum expectation
value would receive a purely non-perturbative correction in that
case irrespective of the space-time dimensions. Hence,
the $\cP\cT$-symmetrically formulated $\lambda\phi^{4}$ theory with
$\lambda<0$ may exhibit a real and positive spectrum, dynamical
symmetry breaking, and asymptotic freedom (and thus non-triviality),
which means in particular that it may be a more suitable candidate
for the Higgs sector in the electroweak theory.
An important lesson here is again the significance
of identifying a linear space on which a given system
shall be considered; a single Hamiltonian (or Lagrangian) can admit
different theories according to different choices of a linear space.

The $\cP\cT$-symmetric formulation has also shed new light on other
quantum field theoretical models. For instance, there have been
some attempts to construct an asymptotically free quantum
electrodynamics with $e^{2}<0$ \cite{BM99,Mi04,BCMS05}.
The controversial Lee model \cite{Le54} was reconsidered from
a viewpoint of $\cP\cT$ symmetry \cite{BBCW05} (see also
Ref.~\cite{Kl04} for some relations between the Lee model and
$\cP\cT$-symmetric theory with the extensive references). There
are also several non-Hermitian models to which the $\cP\cT$-symmetric
approach may apply, from older models such as the $\rmi\varphi^{3}$
theory \cite{Fi78,Ca85} associated with the Lee--Yang edge singularity
\cite{LY52a,LY52b} to newer models such as the timelike Liouville
theory \cite{GS03,ST03}.

However, the $\cP\cT$-symmetric formulation has not yet reached
the level of a physical quantum theory; the emergence of an
indefinite metric has been one of the obstacles.
Toward the construction of a physical theory, there have been
roughly two different approaches, namely, $\cP\cT$-symmetric
Hamiltonians with $\cC$ operators \cite{BBJ02,BBJ04a} and
pseudo-Hermitian Hamiltonians with positive metric operators
\cite{Mo02a,Mo02c}. For a brief description of the development
in the field, see Ref.~\cite{Ta06b}.
The recent progress in the last few years (e.g. Ref.~\cite{MB04})
indicates that both of the approaches are likely to resolve
themselves into classes of the quasi-Hermitian theory proposed
in Ref.~\cite{SGH92}, although some disputes have still persisted,
e.g., Refs.~\cite{BCM06,Mo06a}.
Besides the disputes, both of them have not so far overcome
the following drawbacks sufficiently:
\begin{enumerate}
\item the lack of a systematic prescription independent from
 domains of operators (see e.g. Ref.~\cite{Mo05a} for
 a case-by-case treatment);
\item the lack of a framework applicable even when $\cP\cT$ symmetry
 is spontaneously broken, which is serious since ascertaining
 rigorously unbroken $\cP\cT$ symmetry is extremely difficult
 (see a rigorous proof in Ref.~\cite{DDT01a});
\item unboundedness of $\cC$ or metric operators (see
 Ref.~\cite{KS04}).
\end{enumerate}
In our previous short letter \cite{Ta06b}, we proposed a unified
mathematical framework for $\cP\cT$-symmetric quantum theory
defined in a Krein space, which would be able to surmount the above
difficulties. There by `unified' we meant that its applicability
does not rely on whether a theory is defined on $\bbR$ or a complex
contour, on whether $\cP\cT$ symmetry is unbroken, and so on,
thus it is free from the first and second defects described above.
In the context of $\cP\cT$ symmetry, a Krein space was first
introduced in Ref.~\cite{Ja02} and was then employed in e.g.
Refs.~\cite{LT04,GSZ05}. The Krein space in Ref.~\cite{Ta06b}
can be regarded as a generalization of them. Furthermore,
our framework can circumvent the third difficulty since it is
formulated with a Hilbert space from the beginning and
in this sense we need neither another Hilbert space
nor any metric operator. We also clarified in particular
the relation between $\cP\cT$ symmetry and pseudo-Hermiticity
in our framework. However, it has been still nothing more than
a mathematical framework; no physics has been involved in it.

In this paper, we therefore explore and discuss various general
consequences of $\cP\cT$-symmetric quantum theory defined in
the Krein space, putting our emphasis on whether the theory can be
acceptable as a physical theory. To this end, we examine not only
spectral property and $\cP\cT$ symmetry breaking but also several
issues, crucial for the theory to be physically acceptable,
such as time evolution of state vectors, probability interpretation,
uncertainty relation, classical-quantum correspondence, completeness,
existence of a basis, and so on. We find that the several
significant properties in ordinary quantum theory can also hold
in our case analogously. In particular, we show that
for a given real classical system we can always construct the
corresponding $\cP\cT$-symmetric quantum system, which indicates
that $\cP\cT$-symmetric theory in the Krein space is another
quantization scheme rather than a generalization of the traditional
Hermitian one in the Hilbert space.

We organize the paper as follows. In the next section, we review
the mathematical framework, developed in our previous paper
\cite{Ta06b}, with which we shall discuss various aspects of
$\cP\cT$-symmetric quantum theory in this paper.
Section~\ref{sec:spec} is devoted to spectral properties and
structure of root and eigenspaces in connection with spontaneous
$\cP\cT$ symmetry breaking. In Section~\ref{sec:time}, we
investigate the time evolution of state vectors of
$\cP\cT$-symmetric Hamiltonians and derive a conserved quantity
in time. We then discuss possible ways to a probability
interpretation of matrix elements and derive a couple of criteria
for $\cP\cT$-symmetric Hamiltonians to be physically acceptable.
In Section~\ref{sec:uncer}, we derive an uncertainty
relation hold in $\cP\cT$-symmetric quantum theory.
In Section~\ref{sec:cqco}, we discuss classical-quantum
correspondence in $\cP\cT$-symmetric theory. We first derive
an alternative to the Ehrenfest's theorem in our case, then
discuss its consequences, especially a novel relation to real
classical systems. In Section~\ref{sec:KH}, we access the problem
on completeness and existence of a basis, both of which are
inevitable for the theory to be physically acceptable. We show
that these requirements together with the criteria derived in
Section~\ref{sec:time} naturally restrict operators to the class
$\bK(\bH)$. Finally, we summarize and discuss the results and
propose a postulate for an operator to be a physical observable
in Section~\ref{sec:disc}.

\section{$\cP\cT$-Symmetric Operators in a {K}rein Space}

\subsection{Preliminaries}
\label{ssec:pre}

To begin with, let us introduce a complex-valued smooth function
$\zeta(x)$ on the real line $\zeta :\bbR\to\bbC$ satisfying that
(i) the real part of $\zeta(x)$ is monotone increasing in $x$ such
that $\Re\zeta'(x)>c(>0)$ for all $x\in\bbR$ and thus
$\Re\zeta(x)\to\pm\infty$ as $x\to\pm\infty$, (ii) the first derivative
is bounded, i.e., $(0<)|\zeta'(x)|<C(<\infty)$ for all $x\in\bbR$, and
(iii) $\zeta(-x)=-\zeta^{\ast}(x)$ where $\ast$ denotes complex
conjugate. The function $\zeta(x)$ defines a complex contour in the
complex plane and here we are interested in a family of the following
complex contours:
\begin{align}
\Gamma_{a}\equiv\bigl\{\zeta(x)\bigm|x\in(-a,a),\ a>0\bigr\},
\label{eq:dG}
\end{align}
which has mirror symmetry with respect to the imaginary axis. This
family of complex contours would sufficiently cover all the support
needed to define $\cP\cT$-symmetric quantum mechanical systems.
In particular, we note that $\Gaminf$ with $\zeta(x)=x$ is just
the real line $\bbR$ on which standard quantum mechanical systems
are considered.

Next, we consider a complex vector space $\fF$ of a certain class
of complex functions and introduce a sesquilinear Hermitian form
$Q_{\Gamma_{a}}(\cdot,\cdot):\fF\times\fF\to\bbC$ on the space
$\fF$, with a given $\zeta(x)$, by
\begin{align}
Q_{\Gamma_{a}}(\phi,\psi)\equiv\int_{-a}^{a}\rmd x\,
 \phi^{\ast}(\zeta(x))\psi(\zeta(x)).
\label{eq:inner}
\end{align}
Apparently, it is positive definite, $Q_{\Gamma_{a}}(\phi,\phi)>0$
unless $\phi=0$, and thus defines an inner product on the space $\fF$.
With this inner product we define a class of complex functions
which satisfy
\begin{align}
\lim_{a\to\infty}Q_{\Gamma_{a}}(\phi,\phi)<\infty,
\end{align}
that is, the class of complex functions which are square integrable
(in the Lebesgue sense) in the complex contour $\Gaminf$ with respect
to the \emph{real} integral measure $\rmd x$. We note that this class
is identical with the class of complex functions which are square
integrable in $\Gaminf$ with respect to the complex measure $\rmd z$
along $\Gaminf$. To see this, suppose first $\phi(z)$ belongs to
the former class. Then we have
\begin{align}
\left|\int_{\Gamma_{a}}\rmd z\,|\phi(z)|^{2}\right|\leq\int_{-a}^{a}
 \rmd x\,|\zeta'(x)||\phi(\zeta(x))|^{2}<C\int_{-a}^{a}\rmd x\,
 |\phi(\zeta(x))|^{2}<\infty,
\label{eq:cfsp1}
\end{align}
where we use the property (ii) of the function $\zeta(x)$, and thus
\begin{align}
\left|\int_{\Gaminf}\rmd z\,|\phi(z)|^{2}\right|\leq C
 \lim_{a\to\infty}Q_{\Gamma_{a}}(\phi,\phi)<\infty,
\end{align}
that is, $\phi(z)$ also belongs to the latter. Conversely, if
$\phi(z)$ belongs to the latter class, then we have
\begin{align}
\int_{-a}^{a}\rmd x\,|\phi(\zeta(x))|^{2}<c^{-1}\int_{-a}^{a}\rmd x\,
 \Re\zeta'(x)|\phi(\zeta(x))|^{2}\leq c^{-1}\left|\int_{-a}^{a}
 \rmd x\,\zeta'(x)|\phi(\zeta(x))|^{2}\right|,
\end{align}
where we use the property (i) of the function $\zeta(x)$, and thus
\begin{align}
\lim_{a\to\infty}Q_{\Gamma_{a}}(\phi,\phi)\leq c^{-1}\left|
 \int_{\Gaminf}\rmd z\,|\phi(z)|^{2}\right|<\infty,
\label{eq:cfsp4}
\end{align}
that is, $\phi(z)$ also belongs to the former.

As in the case of $L^{2}(\bbR)$, we can show that this class of
complex functions also constitutes a Hilbert space equipped with
the inner product
\begin{align}
Q_{\Gaminf}(\phi,\psi)\equiv\lim_{a\to\infty}
 Q_{\Gamma_{a}}(\phi,\psi),
\end{align}
which is hereafter denoted by $L^{2}(\Gaminf)$. A Hilbert space
$L^{2}(\Gamma_{a})$ for a finite positive $a$ can be easily defined
by imposing a proper boundary condition at $x=\pm a$.

Before entering into the main subject, we shall define another
concept for later purposes. For a linear differential operator $A$
acting on a linear function space of a variable $x$,
\begin{align}
A=\sum_{n}\alpha_{n}(x)\frac{\rmd^{n}}{\rmd x^{n}},
\label{eq:ldfop}
\end{align}
the transposition $A^{t}$ of the operator $A$ is defined by
\begin{align}
A^{t}=\sum_{n}(-1)^{n}\frac{\rmd^{n}}{\rmd x^{n}}\alpha_{n}(x).
\end{align}
An operator $L$ is said to have \emph{transposition symmetry} if
$L^{t}=L$. If $A$ acts in a Hilbert space $L^{2}(\Gaminf)$, namely,
$A:L^{2}(\Gaminf)\to L^{2}(\Gaminf)$ the following relation holds
for all $\phi(z),\psi(z)\in\fD(A)\cap\fD(A^{t})\subset L^{2}(\Gaminf)$:
\begin{align}
\limaint\rmd x\,\phi(\zeta(x))A^{t}\psi(\zeta(x))=\limaint
 \rmd x\, [A\phi(\zeta(x))]\psi(\zeta(x)).
\label{eq:trans}
\end{align}

\subsection{$\cP$-Metric and a {K}rein Space}
\label{ssec:pmet}

With these preliminaries, we now introduce the linear parity operator
$\cP$ which performs spatial reflection $x\to -x$ when it acts on
a function of a real spatial variable $x$ as
\begin{align}
\cP f(x)=f(-x).
\end{align}
We then define another sesquilinear form $Q_{\Gamma_{a}}
(\cdot,\cdot)_{\cP}:\fF\times\fF\to\bbC$ by
\begin{align}
Q_{\Gamma_{a}}(\phi,\psi)_{\cP}\equiv Q_{\Gamma_{a}}(\cP\phi,\psi).
\label{eq:Pmetric0}
\end{align}
We easily see that this new sesquilinear form is also Hermitian since
\begin{align}
Q_{\Gamma_{a}}(\psi,\phi)_{\cP}&=
 \int_{-a}^{a}\rmd x\,[\cP\psi(\zeta(x))]^{\ast}\phi(\zeta(x))
 =\int_{-a}^{a}\rmd x\,\psi^{\ast}(-\zeta^{\ast}(x))\phi(\zeta(x))
 \notag\\
&=\int_{-a}^{a}\rmd x'\,\psi^{\ast}(-\zeta^{\ast}(-x'))
 \phi(\zeta(-x'))
 =\int_{-a}^{a}\rmd x'\,\psi^{\ast}(\zeta(x'))\cP\phi(\zeta(x'))
 \notag\\
&=Q_{\Gamma_{a}}(\psi,\cP\phi)=Q_{\Gamma_{a}}^{\ast}(\cP\phi,\psi)
 =Q_{\Gamma_{a}}^{\ast}(\phi,\psi)_{\cP},
\label{eq:HsPm}
\end{align}
where we use the Hermiticity of the form (\ref{eq:inner}) as
well as the property (iii). However, it is evident that the form
(\ref{eq:Pmetric0}) is no longer positive definite in general. We call
the indefinite sesquilinear Hermitian form (\ref{eq:Pmetric0})
\emph{$\cP$-metric}.

We are now in a position to introduce the $\cP$-metric into
the Hilbert space $L^{2}(\Gaminf)$. For all $\phi(z),\psi(z)\in
L^{2}(\Gaminf)$ it is given by
\begin{align}
Q_{\Gaminf}(\phi,\psi)_{\cP}\equiv
 \lim_{a\to\infty}Q_{\Gamma_{a}}(\phi,\psi)_{\cP}
 =\lim_{a\to\infty}\int_{-a}^{a}\rmd x\,\phi^{\ast}(-\zeta^{\ast}(x))
 \psi(\zeta(x)).
\label{eq:Pmetric1}
\end{align}
It should be noted that we cannot take the two limits of the integral
bounds, $a\to\infty$ and $-a\to -\infty$, independently in order to
maintain the Hermiticity of the form given in Eq.~(\ref{eq:HsPm}).
Hence, the symbol $\int_{-\infty}^{\infty}\rmd x$ hereafter employed
in this paper is always understood in the following sense:
\begin{align}
\int_{-\infty}^{\infty}\rmd x\,f(x)\equiv\lim_{a\to\infty}
 \int_{-a}^{a}\rmd x\,f(x).
\end{align}
From the definition of $\cP$ and the relation (\ref{eq:HsPm}),
we easily see that the linear operator $\cP$ satisfies $\cP^{-1}=
\cP^{\dagger}=\cP$, where $\dagger$ denotes the adjoint with respect
to the inner product $Q_{\Gaminf}(\cdot,\cdot)$, and thus is a
\emph{canonical} (or \emph{fundamental}) \emph{symmetry} in
the Hilbert space $L^{2}(\Gaminf)$ (Definition 1.3.8 in \cite{AI89}).
Hence, the $\cP$-metric turns to belong to the class of $J$-metric and
the Hilbert space $L^{2}(\Gaminf)$ equipped with the $\cP$-metric
$Q_{\Gaminf}(\cdot,\cdot)_{\cP}$ is a Krein space, which is hereafter
denoted by $L_{\cP}^{2}(\Gaminf)$. Similarly, a Hilbert space
$L^{2}(\Gamma_{a})$ with $Q_{\Gamma_{a}}(\cdot,\cdot)_{\cP}$ is also
a Krein space $L_{\cP}^{2}(\Gamma_{a})$.

A generalization of the framework to many-body systems (described
by $M$ spatial variables $x_{i}$) would be straightforward by
introducing $M$ complex-valued functions $\zeta_{i}(x_{i})$ which
satisfy similar properties of (i)--(iii) with respect to each variable
$x_{i}$ ($i=1,\dots,M$). An inner product on a vector space $\fF$ of
complex functions of $M$ variables is introduced by
\begin{align}
Q_{\Gamma_{\{a_{i}\}}^{M}}(\phi,\psi)=\int_{-a_{1}}^{a_{1}}\rmd x_{1}
 \cdots\int_{-a_{M}}^{a_{M}}\rmd x_{M}\,\phi^{\ast}(\zeta_{1}(x_{1}),
 \dots,\zeta_{M}(x_{M}))\psi(\zeta_{1}(x_{1}),\dots,\zeta_{M}(x_{M})),
\end{align}
where $\Gamma_{\{a_{i}\}}^{M}=\Gamma_{a_{1}}\times\dots\times
\Gamma_{a_{M}}$ with each $\Gamma_{a_{i}}$ given by Eq.~(\ref{eq:dG}).
Then, we can easily follow a similar procedure in the previous and
this subsections to construct a Hilbert space $L^{2}$ and a Krein
space $L_{\cP}^{2}$ in the case of many-body systems.

A canonical decomposition of the Krein space $L_{\cP}^{2}$ is easily
obtained by introducing the canonical orthoprojectors
\begin{align}
P^{\pm}=\frac{1}{2}(I\pm\cP).
\label{eq:P+-}
\end{align}
With them we have
\begin{align}
L_{\cP}^{2}=L_{\cP+}^{2}[\dotplus]L_{\cP-}^{2},\qquad
 L_{\cP\pm}^{2}\equiv P^{\pm}L_{\cP}^{2}=\{\phi\in L_{\cP}^{2}
 |\cP\phi=\pm\phi\},
\label{eq:L+-}
\end{align}
where $[\dotplus]$ denotes $\cP$-orthogonal direct sum.
That is, the positive (negative) subspace $L_{\cP+}^{2}$
($L_{\cP-}^{2}$) is composed of all the $\cP$-even ($\cP$-odd)
vectors in $L_{\cP}^{2}$, respectively.

\subsection{$\cP$-{H}ermiticity and $\cP\cT$ Symmetry}
\label{ssec:pHerm}

Let us next consider a linear operator $A$ acting in the Krein space
$L_{\cP}^{2}$, namely, $A:\fD(A)\subset L_{\cP}^{2}\to\fR(A)\subset
L_{\cP}^{2}$ with non-trivial $\fD(A)$ and $\fR(A)$. The $\cP$-adjoint
of the operator $A$ is such an operator $A^{c}$ that satisfies for all
$\phi\in\fD(A)$
\begin{align}
Q_{\Gaminf}(\phi,A^{c}\psi)_{\cP}=Q_{\Gaminf}(A\phi,\psi)_{\cP},
 \qquad\psi\in \fD(A^{c}),
\label{eq:Padj}
\end{align}
where the domain $\fD(A^{c})$ of $A^{c}$ is determined by the
existence of $A^{c}\psi\in L_{\cP}^{2}$. By the definitions
(\ref{eq:Pmetric1}) and (\ref{eq:Padj}) the $\cP$-adjoint operator
$A^{c}$ satisfies
\begin{align}
Q_{\Gaminf}(\phi,A^{c}\psi)_{\cP}=Q_{\Gaminf}(\cP A\phi,\psi)
 =Q_{\Gaminf}(\phi,A^{\dagger}\cP\psi)
 =Q_{\Gaminf}(\phi,\cP A^{\dagger}\cP\psi)_{\cP},
\end{align}
that is, it is related to the adjoint operator $A^{\dagger}$ in
the corresponding Hilbert space $L^{2}$ by
\begin{align}
A^{c}=\cP A^{\dagger}\cP,\qquad\fD(A^{c})=\fD(A^{\dagger}).
\label{eq:AcAd}
\end{align}
A linear operator $A$ is called \emph{$\cP$-Hermitian} if
$A^{c}=A$ in $\fD(A)\subset L_{\cP}^{2}$, and is called
\emph{$\cP$-self-adjoint} if $\overline{\fD(A)}=L_{\cP}^{2}$ and
$A^{c}=A$.
Here we note that the concept of \emph{$\eta$-pseudo-Hermiticity}
introduced in Ref.~\cite{Mo02a} is essentially equivalent
to what the mathematicians have long called \emph{$G$-Hermiticity}
(with $G=\eta$) among the numerous related concepts in the field
(cf. Sections {1.6} and {2.3} in \cite{AI89}).
Therefore, in this article we exclusively employ the latter
mathematicians' terminology to avoid confusion. Unless specifically
stated, we follow the terminology after the book~\cite{AI89}
supplemented by the one employed in the book~\cite{Bo74}.\footnote{
Some of the terms are different between them. In particular, the
notations and terms regarding the relation between a positive
definite Hilbert space inner product and indefinite metrics are
opposite.}

We now consider so-called $\cP\cT$-symmetric operators in
the Krein space $L_{\cP}^{2}$. The action of the anti-linear
time-reversal operator $\cT$ on a function of a real spatial
variable $x$ is defined by
\begin{align}
\cT f(x)=f^{\ast}(x),
\end{align}
and thus $\cT^{2}=1$ and $\cP\cT=\cT\cP$ follow. Then an
operator $A$ acting on a linear function space $\fF$ is said
to be \emph{$\cP\cT$-symmetric} if it commutes with
$\cP\cT$:\footnote{There exists the notion of \emph{$G$-symmetry}
which lies in an intermediate position between $G$-Hermiticity
and $G$-self-adjointness (cf. Definitions {2.3.1} and {2.3.2} in
\cite{AI89}). But $\cP\cT$ is anti-linear and is not a Gram
operator (cf. Definition {1.6.3} in \cite{AI89}). Thus confusion
would not arise.}
\begin{align}
[\cP\cT, A]=\cP\cT A-A\,\cP\cT=0.
\label{eq:defPT}
\end{align}

To investigate the property of $\cP\cT$-symmetric operators in the
Krein space $L_{\cP}^{2}$, we first note that the $\cP$-metric
can be expressed as
\begin{align}
Q_{\Gamma_{a}}(\phi,\psi)_{\cP}=\int_{-a}^{a}\rmd x\,[\cP
 \phi(\zeta(x))]^{\ast}\psi(\zeta(x))=\int_{-a}^{a}\rmd x\,[\cP\cT
 \phi(\zeta(x))]\psi(\zeta(x)).
\label{eq:Pmetric2}
\end{align}
It is similar to but is slightly different from the (indefinite)
$\cP\cT$ inner product in Ref.~\cite{BBJ02}. Furthermore,
if $\zeta(x)=x$ with finite $a$ or $a\to\infty$, it reduces to
the one considered in Refs.~\cite{Ja02,BQZ01,LT04,GSZ05} and
is essentially equivalent to the indefinite metric introduced
(without the notion of $\cP\cT$) by Pauli in 1943~\cite{Pa43}.

Let $A$ be a $\cP\cT$-symmetric operator in the Krein space
$L_{\cP}^{2}$. By the definitions (\ref{eq:Padj}) and
(\ref{eq:defPT}), and Eqs.~(\ref{eq:trans}) and (\ref{eq:Pmetric2}),
the $\cP$-adjoint of $A$ reads
\begin{align}
Q_{\Gaminf}(\phi,A^{c}\psi)_{\cP}&=\int_{-\infty}^{\infty}\rmd x\,
 [\cP\cT A\phi(\zeta(x))]\psi(\zeta(x))=\int_{-\infty}^{\infty}
 \rmd x\,[A\,\cP\cT\phi(\zeta(x))]\psi(\zeta(x))\notag\\
&=\int_{-\infty}^{\infty}\rmd x\,[\cP\cT\phi(\zeta(x))]A^{t}
 \psi(\zeta(x))=Q_{\Gaminf}(\phi,A^{t}\psi)_{\cP},
\label{eq:PTadj}
\end{align}
that is, $A^{c}=A^{t}$ in $\fD(A^{c})$
for an arbitrary $\cP\cT$-symmetric operator $A$. Hence,
a $\cP\cT$-symmetric operator is $\cP$-Hermitian in $L_{\cP}^{2}$ if
and only if it has transposition symmetry as well. In particular,
since any Schr\"odinger operator $H=-\rmd^{2}/\rmd x^{2}+V(x)$ has
transposition symmetry, $\cP\cT$-symmetric Schr\"odinger operators are
always $\cP$-Hermitian in $L_{\cP}^{2}$. The latter fact naturally
explains the characteristic properties of the $\cP\cT$-symmetric
quantum systems found in the literature; indeed they are completely
consistent with the well-established mathematical consequences of
$J$-Hermitian (more precisely, $J$-self-adjoint) operators in
a Krein space \cite{AI89} with $J=\cP$. Therefore, we can naturally
consider any $\cP\cT$-symmetric quantum system in the Krein space
$L_{\cP}^{2}$, regardless of whether the support $\Gaminf$ is $\bbR$
or not, and of whether $\cP\cT$ symmetry is spontaneously broken or
not. It should be noted, however, that the relation between $\cP\cT$
symmetry and $J$-Hermiticity (more generally $G$-Hermiticity) varies
according to in what kind of Hilbert space we consider operators.
This is due to the different characters of the two concepts; any kind
of Hermiticity is defined in terms of a given inner product while
$\cP\cT$ symmetry is not \cite{GT06}.

Finally, we note that it would be to some extent restrictive to
consider only operators with transposition symmetry although
we are mostly interested in Schr\"odinger operators. For
operators without transposition symmetry, $\cP\cT$ symmetry
does not guarantee $\cP$-Hermiticity. Hence, the requirement
of $\cP\cT$ symmetry alone would be less restrictive as
an alternative to the postulate of self-adjointness in ordinary
quantum mechanics. Furthermore, as we will see later on, even
the stronger condition of $\cP$-self-adjointness turns to be
unsatisfactory from the viewpoint of physical requirements.

\section{Spectral Properties and $\cP\cT$ Symmetry Breaking}
\label{sec:spec}

In this section, we first review some significant mathematical
properties regarding eigenvectors and spectrum of $J$-Hermitian
operators, and then discuss $\cP\cT$ symmetry breaking.
For this purpose, let us first summarize the mathematical definitions
which are indispensable for understanding the characteristic features
of the spectral properties in indefinite metric spaces.

Let $\lambda$ be an eigenvalue of a linear operator $A$ in a linear
space $\fF$, namely, $\lambda\in\sigma_{p}(A)$. The vector
$\phi(\neq 0)$ is a \emph{root} (or \emph{principal}) vector of $A$
belonging to $\lambda$ if there is a natural number $n$ such that
$\phi\in\fD(A^{n})$ and $(A-\lambda I)^{n}\phi=0$. The span of all
the root vectors of $A$ belonging to $\lambda$ is the root subspace
denoted by $\fS_{\lambda}(A)$, namely,
\begin{align}
\fS_{\lambda}(A)=\bigcup_{n=0}^{\infty}
 \Ker\bigl((A-\lambda I)^{n}\bigr).
\label{eq:decom}
\end{align}
The algebraic and geometric multiplicities of $\lambda$, denoted by
$m_{\lambda}^{(a)}(A)$ and $m_{\lambda}^{(g)}(A)$ respectively, are
defined by
\begin{align}
m_{\lambda}^{(a)}(T)=\dim\fS_{\lambda}(A),\qquad
 m_{\lambda}^{(g)}(T)=\dim\Ker(A-\lambda I).
\end{align}
It is evident that $m_{\lambda}^{(a)}\geq m_{\lambda}^{(g)}$ for all
$\lambda$. An eigenvalue $\lambda$ is called \emph{semi-simple} if
$m_{\lambda}^{(a)}=m_{\lambda}^{(g)}$, that is, if $\fS_{\lambda}(A)
=\Ker(A-\lambda I)$. Furthermore, a (semi-simple) eigenvalue $\lambda$
is called \emph{simple} if $m_{\lambda}^{(a)}(=m_{\lambda}^{(g)})=1$.

Let $\fH_{J}$ be a Krein space equipped with a $J$-metric
$Q(\cdot,\cdot)_{J}$. Vectors $\phi,\psi\in\fH_{J}$ are said to be
\emph{$J$-orthogonal} and denoted by $\phi[\perp]\psi$ if
$Q(\phi,\psi)_{J}=0$. Similarly, subspaces
$\fL_{1},\fL_{2}\subset\fH_{J}$ are said to be $J$-orthogonal and
denoted by $\fL_{1}[\perp]\fL_{2}$ if $\phi[\perp]\psi$ for all
$\phi\in\fL_{1}$ and $\psi\in\fL_{2}$. The \emph{$J$-orthogonal
complement} of a set $\fL\subset\fH_{J}$ is the subspace
$\fL^{[\perp]}\subset\fH_{J}$ defined by
\begin{align}
\fL^{[\perp]}=\bigl\{\psi\in\fH_{J}\bigm|\psi[\perp]\fL\bigr\}.
\end{align}
A vector $\phi\in\fH_{J}$ is said to be \emph{neutral} if
$\phi[\perp]\phi$. Similarly, a subspace $\fL$ is said to be neutral
if $\phi[\perp]\phi$ for all $\phi\in\fL$. It follows from the
Cauchy--Schwarz inequality (cf. Eq.~(\ref{eq:CS})), which holds
in any neutral space, that every neutral subspace $\fL$ is
$J$-orthogonal to itself and thus $\fL\subset\fL^{[\perp]}$
(cf. Proposition {1.4.17} in \cite{AI89}).
The \emph{isotropic part} $\fL_{0}$ of a subspace $\fL\subset\fH_{J}$
is defined by $\fL_{0}=\fL\cap\fL^{[\perp]}$ and its (non-zero)
elements are called \emph{isotropic vectors} of $\fL$. In other words,
$\psi_{0}\in\fL$ is an isotropic vector of $\fL$ if
$(0\neq)\psi_{0}[\perp]\fL$. It is evident that the isotropic part
of any subspace is neutral. A subspace $\fL$ is said to be
\emph{non-degenerate} if its isotropic part is trivial,
$\fL_{0}=\{0\}$. Otherwise, it is called \emph{degenerate}.

With these preliminaries, let us review some relevant mathematical
theorems on the structure of root subspaces of a $J$-self-adjoint
operator. One of the most notable ones in our context is
the following (Theorem {II.3.3} in \cite{Bo74}):
\begin{The}
Let $A$ be a $J$-Hermitian operator. If $\lambda$ and $\mu$ are
eigenvalues of $A$ such that $\lambda\neq\mu^{\ast}$, then
$\fS_{\lambda}(A)[\perp]\fS_{\mu}(A)$.
\label{th:Jorth}
\end{The}
The $\cP$-orthogonal relations on $\bbR$ found in
Refs.~\cite{Ja02,BQZ01} are just a special case of
Theorem~\ref{th:Jorth} when both the eigenvalues $\lambda$ and
$\mu$ are semi-simple.
As a consequence of Theorem~\ref{th:Jorth}, we have,
\begin{Cor}
Any root subspace belonging to a non-real eigenvalue of a
$J$-Hermitian operator is neutral.
\label{th:neut}
\end{Cor}
However, it does not guarantee that every root vector belonging to
a real eigenvalue of a $J$-Hermitian operator is non-degenerate.
To see this, suppose $\lambda$ is a non-semi-simple real
eigenvalue of a $J$-Hermitian operator $A$, and let
$Q(\cdot,\cdot)_{J}$ be the $J$-metric. Then there exists
a natural number $n\geq 2$ and a vector $\phi_{n}\in\fD(A^{n})$
which satisfies $(A-\lambda I)^{n}\phi_{n}=0$ and
$(A-\lambda I)^{n-1}\phi_{n}\equiv\phi_{1}\neq 0$. By definition
$\phi_{1}$ is an eigenvector belonging to $\lambda$, namely,
$\phi_{1}\in\Ker(A-\lambda I)$. On the other hand, from
the $J$-Hermiticity of $A$ and the reality of $\lambda$ we have,
for all $\psi\in\Ker(A-\lambda I)$
\begin{align}
Q(\phi_{1},\psi)_{J}&=Q((A-\lambda I)^{n-1}\phi_{n},\psi)_{J}\notag\\
&=Q((A-\lambda I)^{n-2}\phi_{n},(A-\lambda I)\psi)_{J}=0.
\end{align}
That is, $\phi_{1}[\perp]\Ker(A-\lambda I)$ and thus $\phi_{1}$ is
an isotropic eigenvector of $\Ker(A-\lambda I)$. Hence we have,
\begin{Pro}
If a real eigenvalue $\lambda$ of a $J$-Hermitian operator is not
semi-simple, the corresponding eigenspace $\Ker(A-\lambda I)$ is
degenerate.
\label{th:nsemi}
\end{Pro}
It is important to note that the existence of isotropic
eigenvectors of $\Ker(A-\lambda I)$ does not immediately imply
the degeneracy of $\fS_{\lambda}(A)$ since there can exist vectors
of $\fS_{\lambda}(A)\setminus\Ker(A-\lambda I)$ which are \emph{not}
$J$-orthogonal to each isotropic vector of $\Ker(A-\lambda I)$.
As a simple example, let us consider a two-dimensional vector space
$\bbC^{2}$ with an ordinary inner product $(\phi,\psi)=a_{1}^{\ast}
b_{1}+a_{2}^{\ast}b_{2}$ for $\phi=(a_{1},a_{2})^{t}$ and
$\psi=(b_{1},b_{2})^{t}$, and let $A$ and $J$ be operators in
$\bbC^{2}$ and $e_{i}$ ($i=1,2$) be a basis of $\bbC^{2}$ as
the followings:
\begin{align}
A=\left(\begin{array}{rr}\lambda & 1\\ 0 & \lambda\end{array}
 \right),\quad
 J=\left(\begin{array}{rr}0 & 1\\ 1 & 0\end{array}\right),\quad
 e_{1}=\left(\begin{array}{r}1\\0\end{array}\right),\quad
 e_{2}=\left(\begin{array}{r}0\\1\end{array}\right),
\end{align}
where $\lambda\in\bbR$.
Then, it is easy to see that $A$ is $J$-Hermitian, $JA^{\dagger}J
=A$, that $\lambda$ is a non-semi-simple real eigenvalue of $A$,
and that $\Ker(A-\lambda I)=\langle e_{1}\rangle$ and
$\fS_{\lambda}(A)=\bbC^{2}$. On the other hand, $(e_{1},Je_{1})=0$
and $(e_{1},Je_{2})=1$, that is, $\Ker(A-\lambda I)$ is neutral
but $\fS_{\lambda}(A)$ is non-degenerate with respect to
the $J$-metric $(\cdot,J\,\cdot)$.

Regarding the non-degeneracy of the root subspaces, the concept of
normality of eigenvalues plays a key role. An eigenvalue $\lambda$
of a closed linear operator $A$ in a Hilbert space $\fH$ is said
to be \emph{normal} if $(0<)m_{\lambda}^{(a)}(A)<\infty$,
$\fH=\fS_{\lambda}(A)\dotplus\fL$ where $\fL$ is closed,
$A\fL\subset\fL$, and $\lambda\in\rho(A|_{\fL})$. Then the following
theorem holds (Theorem {VI.7.5} in \cite{Bo74}):
\begin{The}
Let $A$ be a $J$-self-adjoint operator with $\rho(A)\neq\emptyset$.
If $\lambda$ is a normal eigenvalue of $A$, so is $\lambda^{\ast}$,
and the root subspaces $\fS_{\lambda}(A)$ and $\fS_{\lambda^{\ast}}
(A)$ are \emph{skewly linked} (or \emph{dual companions}), namely,
$\fS_{\lambda}(A)\cap\fS_{\lambda^{\ast}}(A)^{[\perp]}=
\fS_{\lambda}(A)^{[\perp]}\cap\fS_{\lambda^{\ast}}(A)=\{0\}$,
the relation being denoted by
$\fS_{\lambda}(A)\#\fS_{\lambda^{\ast}}(A)$.\footnote{In this case,
the dimensions of $\fS_{\lambda}$ and $\fS_{\lambda^{\ast}}$ are
finite and thus they admit a $J$-biorthogonal basis (cf. Lemma
{1.1.31} in \cite{AI89}). Note, however, that it is different from
the biorthogonal basis employed in, e.g., the Mostafazadeh's
pseudo-Hermitian formulation \cite{Mo02a}; the former is
$J$-biorthogonal with respect to an indefinite $J$-metric while
the latter is biorthogonal with respect to a positive definite
inner product.}
\label{th:skew}
\end{The}
When $\lambda$ in the above is real, the consequence that
$\fS_{\lambda}$ is skewly linked with itself apparently means
that it is non-degenerate. On the other hand, when $\lambda$ is
non-real, we first note that from Corollary~\ref{th:neut}
$\fS_{\lambda^{\ast}}$ is neutral and thus
$\fS_{\lambda^{\ast}}\subset\fS_{\lambda^{\ast}}^{[\perp]}$.
Suppose $\psi=\phi+\varphi$ ($\phi\in\fS_{\lambda}$,
$\varphi\in\fS_{\lambda^{\ast}}$) is an isotropic vector of
$\fS_{\lambda}\dotplus\fS_{\lambda^{\ast}}$. Then, we have
$\psi\in(\fS_{\lambda}\dotplus\fS_{\lambda^{\ast}})^{[\perp]}
\subset\fS_{\lambda^{\ast}}^{[\perp]}$ on one hand and
$\varphi\in\fS_{\lambda^{\ast}}^{[\perp]}$ on the other hand.
Hence, $\phi\in\fS_{\lambda}\cap\fS_{\lambda^{\ast}}^{[\perp]}$,
but the latter subspace is trivial from $\fS_{\lambda}\#
\fS_{\lambda^{\ast}}$ and thus $\phi=0$. As a result,
$\psi=\varphi\in(\fS_{\lambda}\dotplus\fS_{\lambda^{\ast}}
)^{[\perp]}\subset\fS_{\lambda}^{[\perp]}$ and hence
$\varphi\in\fS_{\lambda}^{[\perp]}\cap\fS_{\lambda^{\ast}}$.
But the latter space is again trivial from $\fS_{\lambda}\#
\fS_{\lambda^{\ast}}$ and thus we finally have $\psi=0$,
that is, the isotropic part of $\fS_{\lambda}\dotplus
\fS_{\lambda^{\ast}}$ is trivial (cf. Lemma I.10.1 in \cite{Bo74}).
Summarizing the above results, we have the following:
\begin{Cor}
Let $\lambda$ be a normal eigenvalue of a $J$-self-adjoint operator
$A$ with $\rho(A)\neq\emptyset$. If $\lambda$ is real, then the root
subspace $\fS_{\lambda}(A)$ is non-degenerate. If $\lambda$ is
non-real, then $\fS_{\lambda}(A)\cap\fS_{\lambda^{\ast}}(A)=\{0\}$
and the subspace $\fS_{\lambda}(A)\dotplus\fS_{\lambda^{\ast}}(A)$
is non-degenerate.
\label{th:nond}
\end{Cor}
Since every finite-dimensional non-degenerate subspace $\fL$ of
a Krein space $\fH_{J}$ is \emph{projectively complete}, namely,
$\fL[\dotplus]\fL^{[\perp]}=\fH_{J}$ (Corollary 1.7.18 in
\cite{AI89}), Corollary~\ref{th:nond} implies that for each
normal eigenvalue $\lambda$ we can decompose the space as
$\fH_{J}=\fS_{\lambda}[\dotplus]\fH'_{J}$ when $\lambda\in\bbR$
and as $\fH_{J}=(\fS_{\lambda}\dotplus\fS_{\lambda^{\ast}})
[\dotplus]\fH'_{J}$ when $\lambda\not\in\bbR$. From
Theorem~\ref{th:Jorth} we can proceed with the $J$-orthogonal
decomposition until the remaining subspace contains no
root vectors corresponding to the normal eigenvalues.
Hence we have,
\begin{Pro}
Let $A$ be a $J$-self-adjoint operator in a Krein space $\fH_{J}$
with $\rho(A)\neq\emptyset$, and suppose the spectrum $\sigma(A)$
consists of only normal eigenvalues. Then $\fH_{J}$ admits
the $J$-orthogonal decomposition as
\begin{align}
\fH_{J}=
 \mathop{[\dotplus]}_{\lambda\subset\bbC^{+}}
 \bigl[\fS_{\lambda}(A)\dotplus\fS_{\lambda^{\ast}}(A)\bigr]
 \mathop{[\dotplus]}_{\lambda\subset\bbR}\fS_{\lambda}(A)
 [\dotplus]\fH'_{J},
\label{eq:alln}
\end{align}
where $\bbC^{+}$ is the set of complex numbers $\lambda$ with
$\Im\,\lambda>0$ and $\sigma(A|_{\fH'_{J}})=\emptyset$.
\label{th:alln}
\end{Pro}

Next, we consider spontaneous $\cP\cT$ symmetry breaking. Let $H$
be a $\cP\cT$-symmetric and $\cP$-self-adjoint operator in
$L_{\cP}^{2}$, and let $\phi\in L_{\cP}^{2}$ be an eigenvector
of $H$ belonging to an eigenvalue $\lambda$. It immediately
follows from $\cP\cT$ symmetry of $H$ that
\begin{align}
H\phi=\lambda\phi\quad\Longrightarrow\quad
 H\cP\cT\phi=\lambda^{\ast}\cP\cT\phi.
\end{align}
We note that $\cP\cT\phi\in L_{\cP}^{2}$ since
\begin{align}
Q_{\Gaminf}(\cP\cT\phi,\cP\cT\phi)_{\cP}&=\int_{-\infty}^{\infty}
 \rmd x\, [\cP\cT\cP\cT\phi(\zeta(x))]\cP\cT\phi(\zeta(x))
 \notag\\
&=\int_{-\infty}^{\infty}\rmd x\,\phi(\zeta(x))\cP\cT\phi(\zeta(x))
 =Q_{\Gaminf}(\phi,\phi)_{\cP}.
\label{eq:PTpL}
\end{align}
Thus, $\cP\cT\phi$ is also an eigenvector of $H$ belonging to
the eigenvalue $\lambda^{\ast}$. It is evident that
$\lambda^{\ast}=\lambda$ if $\phi$ is $\cP\cT$-symmetric,
namely, if $\cP\cT\phi\propto\phi$. In particular, every
eigenvector belonging to a simple real eigenvalue
($m_{\lambda}^{(g)}=1$), must be $\cP\cT$-symmetric.
On the other hand, $\lambda^{\ast}\neq\lambda$ implies
$\cP\cT\phi\not\propto\phi$, that is, a system exhibits spontaneous
$\cP\cT$ symmetry breaking whenever a non-real eigenvalue exists.
Then, subtlety can emerge only
when \emph{simultaneously} $\cP\cT\phi\not\propto\phi$ \emph{and}
$\lambda^{\ast}=\lambda$ for a degenerate\footnote{Here we note the
two different meanings of \emph{degenerate} used in this paper;
the one refers to the non-triviality of the isotropic part of
subspaces and the other to the multiplicity of
eigenvalues.} real eigenvalue $\lambda$
($m_{\lambda}^{(g)}>1$).\footnote{It cannot be the case for
Schr\"odinger operators of a single variable by a similar argument
leading to the no-go theorem in ordinary quantum mechanics which
prohibits the existence of spectral degeneracy in one-dimensional
bound-state problems.} But in the latter case we can always choose
the two linearly independent eigenvectors to be $\cP\cT$-symmetric.
In fact, we easily see
\begin{align}
H\psi_{\pm}=\lambda\psi_{\pm},\qquad\cP\cT\psi_{\pm}=\pm\psi_{\pm},
 \qquad\psi_{\pm}\equiv\frac{1}{2}(I\pm\cP\cT)\phi\in L_{\cP}^{2}.
\label{eq:degPT}
\end{align}
Hence, for an arbitrary real eigenvalue we can always have
$\cP\cT$-symmetric eigenvectors irrespective of the existence of
that kind of spectral degeneracy. We note, however, that
the $\cP\cT$-symmetrically chosen eigenvectors (\ref{eq:degPT})
belonging to the same real eigenvalue are not $\cP$-orthogonal
unless $Q_{\Gaminf}(\cP\cT\phi,\phi)_{\cP}\in\bbR$ since
(cf, Eq.~(\ref{eq:PTpL}))
\begin{align}
Q_{\Gaminf}(\psi_{+},\psi_{-})_{\cP}&=\frac{1}{4}
 Q_{\Gaminf}(\phi+\cP\cT\phi,\phi-\cP\cT\phi)_{\cP}\notag\\
&=\frac{1}{2}\Im\, Q_{\Gaminf}(\cP\cT\phi,\phi)_{\cP}
 =\frac{1}{2}\Im\int_{-\infty}^{\infty}\rmd x\,\phi(\zeta(x))^{2}.
\label{eq:p+p-}
\end{align}
Conversely, if we choose the two linearly independent eigenvectors
belonging to the same real eigenvalue to be $\cP$-orthogonal, they
are no longer eigenstates of $\cP\cT$ in general. Hence, we should
say $\cP\cT$ symmetry is \emph{ill-defined} if there is a degenerate
real eigenvalue for which $\cP\cT$ symmetry and $\cP$-orthogonality
of the corresponding eigenvectors are incompatible.
If a $\cP\cT$-symmetric system $A$ has not only a real spectrum
$\sigma(A)\subset\bbR$ but also the entirely well-defined $\cP\cT$
symmetry, that is, all the eigenvectors are $\cP\cT$-symmetric and
$\cP$-orthogonal with each other, we shall say $\cP\cT$ symmetry is
unbroken \emph{in the strong sense}. If, on the other hand, only
the reality of the spectrum is ascertained, we shall say $\cP\cT$
symmetry is unbroken \emph{in the weak sense}.

It is interesting to note that the integral expression in the last of
Eq.~(\ref{eq:p+p-}) is reminiscent of the one appeared in the context
of the non-analyticity condition of eigenvalues in the coupling
constant of the anharmonic oscillator \cite{BW68,BW69,Si70}:
\begin{align}
\int_{-\infty}^{\infty}\rmd x\,\psi(x;E)^{2}=0,
\end{align}
which is also the necessary and sufficient condition of the
non-semi-simpleness of the eigenvalues in the case. As we shall
briefly discuss in what follows, there is indeed an indication
of some relationship. Although the theorems in Ref.~\cite{Si70}
strongly rely on the parity symmetry of the anharmonic
oscillator, they can be generalized to a certain extent
under a more general assumption, suitable for application to our
present case. Let $H$ be a linear differential operator in $L^{2}$
with transposition symmetry, and let $\lambda$ be an eigenvalue
and $\phi(\zeta(x))\in L^{2}$ be the corresponding eigenvector of
$H$. Suppose there is a unique (up to a multiplicative constant)
solution $\varphi(\zeta(x);\Lambda)$ to the equation
\begin{align}
(H-\Lambda)\varphi(\zeta(x);\Lambda)=0,
\label{eq:HLam1}
\end{align}
in a neighborhood $\fN_{\lambda}$ of $\lambda$ such that
$\varphi(\zeta(x);\Lambda)$ is analytic with respect to $\Lambda$
in $\fN_{\lambda}$ and $\varphi(\zeta(x);\lambda)=\phi(\zeta(x))$.
Differentiating Eq.~(\ref{eq:HLam1}) with respect to $\Lambda$ in
$\fN_{\lambda}$ we obtain
\begin{align}
(H-\Lambda)\chi(\zeta(x);\Lambda)=\varphi(\zeta(x);\Lambda),\qquad
 \chi(\zeta(x);\Lambda)\equiv\frac{\del\varphi(\zeta(x);\Lambda)}{
 \del\Lambda}.
\label{eq:HLam2}
\end{align}
Then, if $\chi(\zeta(x);\lambda)\in L^{2}$, the eigenvalue $\lambda$
is not semi-simple on one hand, since by virtue of Eq.~(\ref{eq:HLam2})
\begin{align}
(H-\lambda)^{2}\chi(\zeta(x);\lambda)=(H-\lambda)\phi(\zeta(x))=0,
\end{align}
which means that $\chi(\zeta(x);\lambda)$ is an \emph{associated}
vector (namely, a root vector which is not an eigenvector) belonging
to $\lambda$. On the other hand, if $\chi(\zeta(x);\lambda)\in L^{2}$
we have
\begin{align}
\int_{-\infty}^{\infty}\rmd x\,\phi(\zeta(x))^{2}
&=\int_{-\infty}^{\infty}\rmd x\,\phi(x)(H-\lambda)\chi(\zeta(x);
 \lambda)\notag\\
&=\int_{-\infty}^{\infty}\rmd x\, [(H-\lambda)\phi(\zeta(x))]
 \chi(\zeta(x);\lambda)=0,
\label{eq:0int}
\end{align}
where we use the transposition symmetry of $H$, Eqs.~(\ref{eq:trans})
and (\ref{eq:HLam2}). Thus under the same condition $\chi\in L^{2}$,
Jordan anomalous behavior and the vanishing of the integral
(\ref{eq:0int}) take place simultaneously.

The above analysis suggests that the eigenvectors belonging to
a real eigenvalue $\lambda$ with
$m_{\lambda}^{(a)}>m_{\lambda}^{(g)}>1$ can be simultaneously
$\cP\cT$-symmetric and $\cP$-orthogonal; the neutral eigenvector
$\phi$ in Eq.~(\ref{eq:0int}) belonging to a non-semi-simple
eigenvalue automatically satisfies the $\cP$-orthogonality
$Q_{\Gaminf}(\psi_{+},\psi_{-})=0$ from Eq.~(\ref{eq:p+p-}).
Thus, in this case $\cP\cT$ symmetry may be well-defined and unbroken
even if $\cP\cT\phi\not\propto\phi$. We further note that when
$\lambda$ is a simple real eigenvalue ($m_{\lambda}^{(g)}=1$)
and thus the corresponding eigenvector $\phi$ must be
$\cP\cT$-symmetric $\cP\cT\phi\propto\phi$,
the vanishing of the integral (\ref{eq:0int}) is equivalent to
the neutrality of $\phi$ since
\begin{align}
\int_{-\infty}^{\infty}\rmd x\,\phi(\zeta(x))^{2}
 \propto\int_{-\infty}^{\infty}\rmd x\, [\cP\cT\phi(\zeta(x))]
 \phi(\zeta(x))=Q_{\Gaminf}(\phi,\phi)_{\cP}=0.
\end{align}
Hence, the emergence of a neutral eigenvector in the real sector
of the spectrum does not necessarily mean spontaneous $\cP\cT$
symmetry breaking. Instead, it can imply Jordan anomalous behavior
as we have just discussed (the converse is always true as
Proposition~\ref{th:nsemi} states). This kind of possibility when
the algebraic multiplicity of a real eigenvalue is greater than $1$
was also indicated by the analysis of Stokes multiplier in
Ref.~\cite{Tr05}.

\begin{table}[ht]
\begin{center}
\tabcolsep=10pt
\begin{tabular}{llll}
\hline
Eigenvalue $\lambda$ & Conditions & Root space & $\cP\cT$ symmetry \\
\hline
non-real & & neutral $\fS_{\lambda}$ & broken \\
         & normal, $\rho(A)\neq\emptyset$ &
 non-degenerate $\fS_{\lambda}\dotplus\fS_{\lambda^{\ast}}$ & \\
\hline
real & normal, $\rho(A)\neq\emptyset$
 & non-degenerate $\fS_{\lambda}$ & \\
     & $m_{\lambda}^{(a)}>m_{\lambda}^{(g)}(\geq 1)$
 & degenerate $\Ker(A-\lambda I)$ & possibly unbroken\\
     & $m_{\lambda}^{(a)}=m_{\lambda}^{(g)}>1$
 & & possibly ill-defined \\
     & $m_{\lambda}^{(g)}=1$ & & unbroken \\
\hline
\end{tabular}
\caption{Aspects of root spaces and $\cP\cT$ symmetry relative to the
 (non-)reality of eigenvalues of an arbitrary $\cP\cT$-symmetric
 and $\cP$-self-adjoint operator $A$.}
\label{tb:asp}
\end{center}
\end{table}
Finally, we summarize in Table~\ref{tb:asp} the consequences
regarding the structure of root spaces and $\cP\cT$ symmetry breaking
relative to the (non-)reality of eigenvalues of $\cP\cT$-symmetric
and $\cP$-self-adjoint operators discussed in this section.
In the fourth column of Table~\ref{tb:asp} `possibly' means that
more rigorous case-by-case studies would be needed for ascertaining
the statement.

\section{Time Evolution and Probability Interpretation}
\label{sec:time}

Next, we shall examine the time evolution of quantum state vectors
of a $\cP\cT$-symmetric Hamiltonian. It is determined by the
time-dependent Schr\"odinger equation:
\begin{align}
\rmi\hbar\frac{\del}{\del t}\Psi(\zeta(x),t)=H\Psi(\zeta(x),t)
 =\left[-\frac{\hbar^{2}}{2m}\frac{\del^{2}}{\del x^{2}}+V(x)
 \right]\Psi(\zeta(x),t),
\label{eq:tdSch}
\end{align}
where $\Psi(z,t)\in L_{\cP}^{2}$ and the Hamiltonian $H$ is an
operator acting in $L_{\cP}^{2}$ and satisfying $\cP\cT H=H\cP\cT$.
Here we note a novel feature of our framework. In the conventional
treatment, we consider a Schr\"odinger operator of a complex
variable $z$ along a contour $\Gaminf$ which has mirror symmetry
with respect to the imaginary axis:
\begin{align}
\rmi\hbar\frac{\del}{\del t}\Psi(z,t)=\left[-\frac{\hbar^{2}}{2m}
\frac{\del^{2}}{\del z^{2}}+V(z)\right]\Psi(z,t).
\label{eq:conv1}
\end{align}
Hence, if we parametrize the path as $z=\zeta(x)$ in terms of
a real variable $x$, we have in the conventional approach
\begin{align}
\rmi\hbar\frac{\del}{\del t}\Psi(\zeta(x),t)=\left[
-\frac{\hbar^{2}}{2m}\frac{1}{\zeta'(x)}\frac{\del}{\del x}
\frac{1}{\zeta'(x)}\frac{\del}{\del x}+V(\zeta(x))\right]\Psi(
\zeta(x),t).
\label{eq:conv2}
\end{align}
This difference (when $\zeta(x)\neq x$) is related with the
different choices of a metric. In our Krein space $L_{\cP}^{2}$
the $\cP$-metric is defined with respect to the real measure
$\rmd x$ irrespective of the choice of $\zeta(x)$ and linear
operators we have been considering in the space are differential
operators of a real variable $x$, Eq.~(\ref{eq:ldfop}).
In the conventional treatment, on the other hand, we consider
Schr\"odinger operators of a complex variable $z=\zeta(x)$ and
the naturally induced metric is defined in terms of the complex
measure $\rmd z$, e.g., Refs.~\cite{BBJ02,BBJ04a}.\footnote{
As a set of square integrable complex functions with respect to
each metric they are identical as has been shown in
Eqs.~(\ref{eq:cfsp1})--(\ref{eq:cfsp4}). Hence, every wave function
$\Psi(z,t)$ in the conventional framework (\ref{eq:conv1}) also
belongs to $L_{\cP}^{2}(\Gaminf)$. But we note that the latter
metric is not Hermitian in general.}
Mathematically, this difference is just different settings of
eigenvalue problems. But as we will show, our framework provides
us with a natural way to construct a physically acceptable theory.

Acting the operator $\cP\cT$ to the time-dependent Schr\"odinger
equation (\ref{eq:tdSch}), we have
\begin{align}
-\rmi\hbar\frac{\del}{\del t}\Psi^{\ast}(-\zeta^{\ast}(x),t)
 =H\Psi^{\ast}(-\zeta^{\ast}(x),t)=\left[-\frac{\hbar^{2}}{2m}
 \frac{\del^{2}}{\del x^{2}}+V(x)\right]\Psi^{\ast}(-\zeta^{\ast}(x),
 t),
\label{eq:tdSch'}
\end{align}
where $\cP\cT$ symmetry of $H$ and the property (iii) of $\zeta$
are used.
For a given initial state at $t=t_{0}$, $\Psi(\zeta(x),t_{0})$,
a formal solution to Eq.~(\ref{eq:tdSch}) is given by
\begin{align}
\Psi(\zeta(x),t)=U(t,t_{0})\Psi(\zeta(x),t_{0}),\qquad
 U(t,t_{0})=\rme^{-\rmi H(t-t_{0})/\hbar}.
\label{eq:fsol}
\end{align}
In contrast to ordinary quantum theory, the Hamiltonian $H$ is
non-Hermitian in the Hilbert space $L^{2}$ but is $\cP$-Hermitian
in the Krein space $L_{\cP}^{2}$. Hence, the time evolution operator
$U(t,t_{0})$ is non-unitary in $L^{2}$ but would be
\emph{$\cP$-isometric} (possibly \emph{$\cP$-unitary} when $H$ is
$\cP$-self-adjoint) in $L_{\cP}^{2}$ (cf. Definition {2.5.1} in
\cite{AI89}). To show the latter formally, we first note that for
a $\cP\cT$-symmetric Hamiltonian $H$ we have
\begin{align}
\cP\cT U(t,t_{0})=\cP\cT\sum_{n=0}^{\infty}
 \frac{(-\rmi)^{n}(t-t_{0})^{n}}{\hbar^{n}n!}H^{n}
 =\sum_{n=0}^{\infty}\frac{\rmi^{n}(t-t_{0})^{n}}{\hbar^{n}n!}
 H^{n}\cP\cT=U(t_{0},t)\cP\cT.
\end{align}
The transposition of $U(t,t_{0})$ reads
\begin{align}
U(t,t_{0})^{t}=\biggl(\sum_{n=0}^{\infty}
 \frac{(-\rmi)^{n}(t-t_{0})^{n}}{\hbar^{n}n!}H^{n}\biggr)^{t}
 =\sum_{n=0}^{\infty}\frac{(-\rmi)^{n}(t-t_{0})^{n}}{\hbar^{n}n!}
 H^{n}=U(t,t_{0}),
\end{align}
that is, the time evolution operator has transposition symmetry so
long as the Hamiltonian has. Thus, for an arbitrary
$\Psi=\Psi(\zeta(x),t_{0})\in\fD(U)$ and a time $t$ we obtain
\begin{align}
Q_{\Gaminf}(U\Psi,U\Psi)_{\cP}&=\int_{-\infty}^{\infty}\rmd x
 \left[\cP\cT U(t,t_{0})\Psi(\zeta(x),t_{0})\right]U(t,t_{0})
 \Psi(\zeta(x),t_{0})\notag\\
&=\int_{-\infty}^{\infty}\rmd x\left[ U(t_{0},t)\cP\cT
 \Psi(\zeta(x),t_{0})\right]U(t,t_{0})\Psi(\zeta(x),t_{0})\notag\\
&=\int_{-\infty}^{\infty}\rmd x\left[\cP\cT\Psi(\zeta(x),t_{0})
 \right]U(t_{0},t)U(t,t_{0})\Psi(\zeta(x),t_{0})\notag\\
&=Q_{\Gaminf}(\Psi,\Psi)_{\cP},
\end{align}
which shows the $\cP$-isometric property of $U(t,t_{0})$. Putting
mathematical rigor aside, the above result indicates that we should
replace the requirement of unitarity, of $S$ matrix for instance,
imposed in ordinary quantum theory by that of $\cP$-unitarity
in the case of $\cP\cT$-symmetric quantum theory. For a more
rigorous treatment, namely, a counterpart of the Stone's theorem
in indefinite metric spaces, see Ref.~\cite{Na66}. 

Next, we shall address the issue of the probability interpretation.
{}From the time-dependent Schr\"odinger equation
(\ref{eq:tdSch}) and its $\cP$-adjoint version (\ref{eq:tdSch'}),
the following continuity equation, which is a generalization of
the one derived in Ref.~\cite{BQZ01}, holds for arbitrary solutions
$\Psi_{i}\in L_{\cP}^{2}$ ($i=1,2$) of Eq.~(\ref{eq:tdSch}):
\begin{align}
\frac{\del}{\del t}\bigl[\Psi_{1}^{\ast}(-\zeta^{\ast}(x),t)
 \Psi_{2}(\zeta(x),t)\bigr]=-\frac{\del}{\del x}\cJ(\zeta(x),t),
\label{eq:cont}
\end{align}
where the current density $\cJ$ is defined by
\begin{align}
\cJ(\zeta(x),t)=\frac{\hbar}{2m\rmi}
 \left[\Psi_{1}^{\ast}(-\zeta^{\ast}(x),t)\frac{\del}{\del x}
 \Psi_{2}(\zeta(x),t)-\Psi_{2}(\zeta(x),t)\frac{\del}{\del x}
 \Psi_{1}^{\ast}(-\zeta^{\ast}(x),t)\right].
\end{align}
Integrating both sides of the continuity equation (\ref{eq:cont})
with respect to $x\in(-\infty,\infty)$ we obtain the conservation law
of the $\cP$-metric:
\begin{align}
\frac{\del}{\del t}\int_{-\infty}^{\infty}\rmd x\,\Psi_{1}^{\ast}
 (-\zeta^{\ast}(x),t)\Psi_{2}(\zeta(x),t)=\frac{\del}{\del t}
 Q_{\Gaminf}(\Psi_{1},\Psi_{2})_{\cP}=0.
\label{eq:cons}
\end{align}
This means that though the $\cP$-metric is indefinite, the character
of each state vector, namely, positivity, negativity, or neutrality,
remains unchanged in the time evolution. The conservation law of this
kind is indispensable for the probability interpretation. To examine
further the possibility of it in our framework, we first note that
the emergence of negative norm itself would not immediately mean
the inability of it since probabilities of physical process are
eventually given by the \emph{absolute value} of a certain metric
(matrix element) but not by its complex value itself. Thus, we should
here stress the fact that the true obstacle is the violation of
the Cauchy-Schwarz inequality
\begin{align}
\bigl|Q(\phi,\psi)\bigr|^{2}\leq Q(\phi,\phi) Q(\psi,\psi),
\label{eq:CS}
\end{align}
in indefinite spaces, where $Q(\cdot,\cdot)$ is a sesquilinear
Hermitian form $\fF\times\fF\to\bbC$. The inequality ensures that
the absolute value of an arbitrary matrix element $Q(\phi,\psi)$
is less than or equal to $1$ so long as every vector in $\fF$ is
normalized with respect to $Q(\cdot,\cdot)$, from which we can
assign probability to the quantity $|Q(\phi,\psi)|^{2}$. To see
the violation of the inequality (\ref{eq:CS}) in an indefinite
space, let us consider a two-dimensional vector space $\bbC^{2}$
equipped with an indefinite metric $Q$ defined by
$Q(\phi,\psi)=a_{1}^{\ast}b_{1}-a_{2}^{\ast}b_{2}$ for
$\phi=(a_{1},a_{2})^{t}$ and $\psi=(b_{1},b_{2})^{t}$.
The two vectors $e_{+}=(\sqrt{2},-1)^{t}$ and
$e_{-}=(1,\sqrt{2})^{t}$ are normalized in the sense of
$|Q(e_{\pm},e_{\pm})|=1$, but $|Q(e_{-},e_{+})|=2\sqrt{2}\not\leq 1$.
Hence, the fact that the Cauchy-Schwarz inequality holds at most in
semi-definite spaces (cf. Proposition {1.1.16} in \cite{AI89})
indicates that we must always restrict ourselves to considering
quantum process in a semi-definite subspace in order to make
a probability interpretation in any kind of quantum-like theory with
an indefinite metric.

This observation naturally leads us to consider a pair of
subspaces $(\fL_{+},\fL_{-})$ of the Krein space $L_{\cP}^{2}$
where $\fL_{+}$ (respectively, $\fL_{-}$) is a non-negative
(respectively, a non-positive) subspace of $L_{\cP}^{2}$ and
$\fL_{+}[\perp]\fL_{-}$. This kind of pair corresponds to what is
called a \emph{dual pair} (Definition {1.10.1} in \cite{AI89}).
Then, in each semi-definite subspace
$\fL_{+}$ or $\fL_{-}$, the inequality (\ref{eq:CS}) with
$Q(\cdot,\cdot)=Q_{\Gaminf}(\cdot,\cdot)_{\cP}$ certainly holds.
However, the situation has not been satisfactory yet. To see it,
suppose, at the initial time $t_{0}$, we have a normalized positive
physical state $\Psi_{t_{0}}\equiv\Psi(\zeta(x),t_{0})\in\fL_{+}$.
After the time evolution determined by Eq.~(\ref{eq:fsol}),
the state $\Psi_{t}\equiv\Psi(\zeta(x),t)$ at $t>t_{0}$ remains
positive by virtue of the conservation law (\ref{eq:cons}).
However, it does not necessarily mean $\Psi_{t}\in\fL_{+}$; in
general we have
\begin{align}
\Psi_{t}=\Psi_{t}^{+}+\Psi_{t}^{-},\quad\Psi_{t}^{+}\in\fL_{+},
 \quad \Psi_{t}^{-}\in L_{\cP}^{2}\setminus\fL_{+},\quad
 Q_{\Gaminf}(\Psi_{t},\Psi_{t})_{\cP}>0,
\end{align}
with a non-zero $\Psi_{t}^{-}\in L_{\cP}^{2}\setminus\fL_{+}$.
As a consequence, we cannot consider the matrix element
$Q_{\Gaminf}(\Psi_{t},\Psi_{t_{0}})_{\cP}$ in
the initially prepared positive semi-definite subspace $\fL_{+}$.
This situation would be hardly acceptable since we cannot choose
and fix beforehand a semi-definite subspace where we should consider
a physical process. This difficulty would not arise only when
the Hamiltonian $H$, and thus the time evolution operator $U$,
preserves the initial semi-definite subspaces $\fL_{+}$ and $\fL_{-}$
separately, namely, $\overline{\fD(H)\cap\fL_{\pm}}=\fL_{\pm}$ and
$H(\fD(H)\cap\fL_{\pm})\subset\fL_{\pm}$. If the latter is the case,
the state vector $\Psi_{t}$ at every time $t>t_{0}$ stays in
the semi-definite subspace $\fL_{+}$ or $\fL_{-}$ if the initial
state $\Psi_{t_{0}}$ is an element of $\fL_{+}$ or $\fL_{-}$,
respectively, and it makes sense to regard the quantity
$Q_{\Gaminf}(\Psi_{t},\Psi_{t_{0}})_{\cP}$ in the corresponding
subspace as a transition amplitude, as in ordinary quantum theory.
Furthermore, for arbitrary
$\Psi_{t_{1}}^{+}\equiv\Psi^{+}(\zeta(x),t_{1})\in\fL_{+}$ and
$\Psi_{t_{2}}^{-}\equiv\Psi^{-}(\zeta(x),t_{2})\in\fL_{-}$ we have
$Q_{\Gaminf}(\Psi_{t_{1}}^{+},\Psi_{t_{2}}^{-})_{\cP}=0$, that is,
there is no transition between states in $\fL_{+}$ and $\fL_{-}$.
But both of $\fL_{+}$ and $\fL_{-}$ should be wide enough such that
every element of $L_{\cP}^{2}$ can contribute dynamics. This naturally
leads to the requirement that $\fL_{+}\in\fM^{+}(L_{\cP}^{2})$ and
$\fL_{-}\in\fM^{-}(L_{\cP}^{2})$ where $\fM^{+}(\fH_{J})$
($\fM^{-}(\fH_{J})$) is the set of all \emph{maximal} non-negative
(non-positive) subspaces of a Krein space $\fH_{J}$, respectively.
That is, $P^{\pm}\fL_{\pm}=L_{\cP\pm}^{2}$ where $P^{\pm}$ and
$L_{\cP\pm}^{2}$ are defined by Eqs.~(\ref{eq:P+-}) and
(\ref{eq:L+-}), respectively (cf. Theorem {1.4.5} in \cite{AI89}).
However, operators in a Krein space do not always have such a pair
of invariant maximal semi-definite subspaces, and thus we now arrive
at a criterion for a given Hamiltonian acting in $L_{\cP}^{2}$ to be
physically acceptable:
\begin{description}
 \item[Criterion 1] A $\cP$-self-adjoint Hamiltonian $H$ can be
  physically acceptable if and only if it admits an invariant maximal
  dual pair $(\fL_{+},\fL_{-})$, and thus, so does the one-parameter
  family of $\cP$-unitary operators $U=\rme^{-\rmi tH}$.
\end{description}

An immediate consequence of the inequality (\ref{eq:CS}), which now
holds in each invariant semi-definite subspace $\fL_{\pm}$, is that
every neutral vector $\psi_{\pm}^{(0)}\in\fL_{\pm}$ is isotropic in
each $\fL_{\pm}$, namely, $\psi_{\pm}^{(0)}[\perp]\fL_{\pm}$. Hence,
every transition amplitude with initial or final neutral state is
identically zero. So, it is natural to consider a decomposition
of each semi-definite sector $\fL_{\pm}$ into its isotropic part
$\fL_{\pm}^{0}$ and a definite part $\fL_{\pm\pm}$:
\begin{align}
\fL_{\pm}=\fL_{\pm}^{0}[\dotplus]\fL_{\pm\pm},\qquad
 \fL_{\pm}^{0}=\fL_{\pm}\cap\fL_{\pm}^{[\perp]},
\end{align}
where $\fL_{++}$ ($\fL_{--}$) is a positive (negative) definite
subspace, respectively.
It is now apparent that the $\cP$-metric restricted in each
$\fL_{\pm\pm}$, namely, $\pm Q_{\Gaminf}(\phi,\psi)_{\cP}$
($\phi,\psi\in\fL_{\pm\pm}$) is positive definite. Therefore, we
can finally make a probability interpretation in each definite
subspace $\fL_{\pm\pm}$ separately, provided that every vector
$\psi_{\pm}$ in $\fL_{\pm\pm}$ is normalized with respect to
the \emph{intrinsic $\cP$-norm} $|\,\cdot\,|_{\fL_{\pm\pm}}$ on
$\fL_{\pm\pm}$ such that
\begin{align}
|\psi_{\pm}|_{\fL_{\pm\pm}}\equiv\sqrt{\left|Q_{\Gaminf}(\psi_{\pm},
 \psi_{\pm})_{\cP}\right|}=1.
\label{eq:Pnorm}
\end{align}
Considering mathematical subtleties such as the continuity of the
$\cP$-metric restricted on to $\fL_{\pm\pm}$, we would conclude that
$\fL_{\pm\pm}$ should be \emph{uniformly} definite, which in our
case means the equivalence between the intrinsic $\cP$-norm
$|\psi_{\pm}|_{\fL_{\pm\pm}}$ and the Hilbert space norm $\|\psi_{\pm}
\|\equiv\sqrt{Q_{\Gaminf}(\psi_{\pm},\psi_{\pm})}$ defined in $L^{2}$
for all $\psi_{\pm}\in\fL_{\pm\pm}$, respectively (cf. Section {1.5}
in \cite{AI89}). Hence, we arrive at the second criterion:
\begin{description}
 \item[Criterion 2] Each of the subspaces $\fL_{\pm}$ in Criterion 1
  should admit a decomposition $\fL_{\pm}=\fL_{\pm}^{0}[\dotplus]
  \fL_{\pm\pm}$ into a $\cP$-orthogonal direct sum of its isotropic
  part $\fL_{\pm}^{0}$ and a uniformly definite subspace
  $\fL_{\pm\pm}$, respectively.
\end{description}
The subspaces $\fL_{\pm\pm}$ are then complete relative to
the intrinsic $\cP$-norm $|\,\cdot\,|_{\fL_{\pm\pm}}$, respectively
(Proposition {1.5.6} in \cite{AI89}).
Thus $\fL_{\pm\pm}$ with the positive definite \emph{intrinsic
$\cP$-metric} $\pm Q_{\Gaminf}(\cdot,\cdot)_{\cP}|_{\fL_{\pm\pm}}$
are Hilbert spaces, respectively.
In particular, the time evolution operator in Eq.~(\ref{eq:fsol})
restricted on to $\fL_{\pm\pm}$, $U(t,t_{0})|_{\fL_{\pm\pm}}$ is
unitary in each of the Hilbert spaces $\fL_{\pm\pm}$.

A natural way to construct physical spaces is to consider the
quotient spaces $\tilde{\fL}_{\pm}=\fL_{\pm}/\fL_{\pm}^{0}$ in each
of the sectors. An element $\tilde{\psi}_{\pm}\in\tilde{\fL}_{\pm}$
is defined by the formula $\tilde{\psi}_{\pm}=\psi_{\pm}+\fL_{\pm}^{0}$
for each $\psi_{\pm}\in\fL_{\pm\pm}$, respectively. An induced
positive definite metric $\tilde{Q}_{\pm}(\cdot,\cdot)$ is respectively
given by
\begin{align}
\tilde{Q}_{\pm}(\tilde{\phi}_{\pm},\tilde{\psi}_{\pm})
 =\pm Q_{\Gaminf}(\phi_{\pm},\psi_{\pm})_{\cP},\qquad
 \phi_{\pm},\psi_{\pm}\in\fL_{\pm\pm}.
\end{align}
Then, the quotient space $\tilde{\fL}_{+}$ ($\tilde{\fL}_{-}$) is
isometrically (skew-symmetrically) isomorphic to the positive
(negative) definite subspace $\fL_{++}$ ($\fL_{--}$), respectively
(Proposition {1.1.23} in \cite{AI89}).
This kind of prescription was already employed, e.g., in the BRST
quantization of non-Abelian gauge theories; the whole state vector
space of the latter systems is also indefinite and the positive
definite physical space is given by the quotient space
$\Ker\cQ_{B}/\im\cQ_{B}$, where $\cQ_{B}$ is a nilpotent BRST
charge \cite{BRS76} and $\im\cQ_{B}$ is the BRST-exact neutral
subspace of the BRST-closed non-negative state vector space
$\Ker\cQ_{B}$ \cite{KO78a} (for a review see, e.g.,
Ref.~\cite{KO79}).

Finally, we can classify the set of the systems which satisfy
Criteria 1 and 2 according to the dimension of the subspaces
$\fL_{\pm}^{0}$ and $\fL_{\pm\pm}$:
\begin{description}
\item[Case 1] $\dim\fL_{\pm}^{0}<\infty$
 and $\dim\fL_{\pm\pm}=\infty$, respectively;
\item[Case 2] $\dim\fL_{\pm}^{0}=\infty$
 and $\dim\fL_{\pm\pm}=\infty$, respectively;
\item[Case 3] $\dim\fL_{\pm}^{0}=\infty$
 and $\dim\fL_{\pm\pm}<\infty$, respectively.
\end{description}
In Case 3, the physically relevant space $\fL_{++}$ or $\fL_{--}$
is finite-dimensional and thus the system, at least as physical,
would be less interesting. So, our main concern would be for
systems corresponding to Cases 1 and 2. In connection with Case 1,
we recall a special class of semi-definite subspaces of a Krein
space. A non-negative (non-positive) subspace $\fL$ of a Krein
space $\fH_{J}$ is called a subspace of \emph{class} $h^{+}$
(\emph{class} $h^{-}$) if it admits a decomposition $\fL=\fL^{0}
[\dotplus]\fL^{+}$ ($\fL=\fL^{0}[\dotplus]\fL^{-}$) into a direct
$J$-orthogonal sum of a \emph{finite-dimensional isotropic}
subspace $\fL^{0}$ ($\dim\fL^{0}<\infty$) and a \emph{uniformly
positive} (\emph{uniformly negative}) subspace $\fL^{+}$
($\fL^{-}$)~\cite{Az76}.
We easily see that in Case 1 the semi-definite subspace $\fL_{+}$
($\fL_{-}$) belongs to the class $h^{+}$ ($h^{-}$), respectively.
We will later see in Section~\ref{sec:KH} that there exists a class
of $\cP$-self-adjoint operators which satisfies Criteria 1 and 2
corresponding to Case 1.

\section{Uncertainty Relation}
\label{sec:uncer}

In the previous section, we have derived the criteria for a
quantum-like theory with an indefinite metric to be physically
acceptable from the viewpoint of the probability interpretation.
They are, of course, not sufficient at all. One of the most
crucial criteria is the existence of a secure correspondence
to classical theory. In this and the next sections, we shall discuss
in detail the correspondence between our $\cP\cT$-symmetric theory
in the Krein space $L_{\cP}^{2}$ and classical mechanics. Two
principal issues shall be addressed, namely, uncertainty relation
in this section, and classical equations of motion in the next
section.

The results in the preceding section indicates that the $\cP$-metric
in the Krein space $L_{\cP}^{2}$ rather than the inner product
(\ref{eq:inner}) in the Hilbert space $L^{2}$ plays a central role
in $\cP\cT$-symmetric quantum theory. Hence, we define the
expectation value of an operator $\hat{O}$ in the Krein space
$L_{\cP}^{2}$ by
\begin{align}
\langle\hat{O}\rangle_{\cP}\equiv Q_{\Gaminf}(\Psi,O\Psi)_{\cP}
 =\int_{-\infty}^{\infty}\rmd x\,\Psi^{\ast}(-\zeta^{\ast}(x),t)
 O\Psi(\zeta(x),t),
\label{eq:expv}
\end{align}
where $O$ denotes the $x$-representation of the operator $\hat{O}$
and $\Psi\in L_{\cP}^{2}$ is a solution of the time-dependent
Schr\"odinger equation (\ref{eq:tdSch}). We shall call the quantity
defined by Eq.~(\ref{eq:expv}) \emph{$\cP$-expectation value}.
It reduces to the one considered in e.g. Ref.~\cite{Ja02} when
$\Gaminf=\bbR$.

We first examine $\cP$-adjoint operators relevant in both classical
and quantum theories, namely, scalar multiplication, position, and
momentum operators. From the relation (\ref{eq:AcAd}) we
immediately have
\begin{align}
\lambda^{c}&=\cP\lambda^{\dagger}\cP=\lambda^{\ast}\qquad
 (\lambda\in\bbC),
\label{eq:adjlam}\\
\hat{x}^{c}&=\cP\hat{x}^{\dagger}\cP=-\hat{x},
\label{eq:adjx}\\
\hat{p}^{c}&=\cP\hat{p}^{\dagger}\cP=-\hat{p}.
\label{eq:adjp}
\end{align}
The last two relations show that both the position and momentum
operators are \emph{anti-$\cP$-Hermitian} in the Krein space
$L_{\cP}^{2}$. Since physical quantities are usually expressed as
functions of position and momentum, we shall mainly consider
anti-$\cP$-Hermitian operators as well as $\cP$-Hermitian operators.

It follows from the Hermiticity (\ref{eq:HsPm}) of the $\cP$-metric
that $\cP$-expectation values (\ref{eq:expv}) of
(anti-)$\cP$-Hermitian operators are all real (purely imaginary),
respectively:
\begin{align}
\langle\hat{A}\rangle_{\cP}^{\ast}&=Q_{\Gaminf}^{\ast}
 (\Psi,A\Psi)_{\cP}=Q_{\Gaminf}(A\Psi,\Psi)_{\cP}\notag\\
&=Q_{\Gaminf}(\Psi,A^{c}\Psi)_{\cP}
 =\pm Q_{\Gaminf}(\Psi,A\Psi)_{\cP}\notag\\
&=\pm\langle\hat{A}\rangle_{\cP}\qquad\text{for}\quad\hat{A}^{c}
 =\pm\hat{A}.
\label{eq:ReIm}
\end{align}
Next, we shall consider a commutation relation between two linear
operators
\begin{align}
[\hat{A},\hat{B}]=\hat{A}\hat{B}-\hat{B}\hat{A}=\rmi\hbar\hat{C}.
\label{eq:comAB}
\end{align}
It is easy to see, that the operator $\hat{C}$ is $\cP$-Hermitian
when both $\hat{A}$ and $\hat{B}$ are simultaneously either
$\cP$-Hermitian or anti-$\cP$-Hermitian, and that $\hat{C}$ is
anti-$\cP$-Hermitian when one of $\hat{A}$ and $\hat{B}$ is
$\cP$-Hermitian and the other is anti-$\cP$-Hermitian.

As in the case of ordinary quantum mechanics, we introduce the
\emph{deviation} operator $\Delta\hat{A}$ of $\hat{A}$ by
\begin{align}
\Delta\hat{A}=\langle\hat{A}\rangle_{\cP}-\hat{A}.
\end{align}
The deviation operator $\Delta\hat{A}$ is (anti-)$\cP$-Hermitian
when $\hat{A}$ is (anti-)$\cP$-Hermitian, respectively; for from
Eqs.~(\ref{eq:adjlam}) and (\ref{eq:ReIm}),
\begin{align}
(\Delta\hat{A})^{c}&=\langle\hat{A}\rangle_{\cP}^{\ast}
 -\hat{A}^{c}=\pm\langle\hat{A}\rangle_{\cP}\mp\hat{A}\notag\\
&=\pm\Delta\hat{A}\qquad\text{for}\quad\hat{A}^{c}=\pm\hat{A},
\end{align}
follows. For two operators $\hat{A}$ and $\hat{B}$ satisfying
the commutation relation (\ref{eq:comAB}), the corresponding
deviation operators $\Delta\hat{A}$ and $\Delta\hat{B}$ satisfy
the same relation:
\begin{align}
[\Delta\hat{A},\Delta\hat{B}]=\rmi\hbar\hat{C}.
\label{eq:dAdB}
\end{align}

We are now in a position to discuss uncertainty relation in
$\cP\cT$-symmetric quantum theory. Due to the indefiniteness of
the $\cP$-metric, however, it would be difficult to establish
a certain inequality in the whole Krein space; we recall the violation
of the inequality (\ref{eq:CS}). But the discussion in the previous
section shows that we must always consider physics in a semi-definite
subspace $\fL_{-}$ or $\fL_{+}$. In this sense, we would satisfy
ourselves by establishing an uncertainty relation which holds only
in semi-definite subspaces $\fL_{\pm}$.

Let us suppose that $\hat{A}$ and $\hat{B}$ are respectively either
$\cP$-Hermitian or anti-$\cP$-Hermitian, and that both the vectors
$\Delta A\Psi$ and $\Delta B\Psi$ are simultaneously elements of
a given semi-definite subspace $\fL_{+}$ or $\fL_{-}$. Under these
assumptions, we first note that
\begin{align}
\bigl|Q_{\Gaminf}(\Psi,[\Delta A,\Delta B]\Psi)_{\cP}\bigr|
&\leq\bigl|Q_{\Gaminf}(\Psi,\Delta A\Delta B\Psi)_{\cP}\bigr|
 +\bigl|Q_{\Gaminf}(\Psi,\Delta B\Delta A\Psi)_{\cP}\bigr|\notag\\
&=\bigl|Q_{\Gaminf}(\Delta A\Psi,\Delta B\Psi)_{\cP}\bigr|
 +\bigl|Q_{\Gaminf}(\Delta B\Psi,\Delta A\Psi)_{\cP}\bigr|\notag\\
&=2\bigl|Q_{\Gaminf}(\Delta A\Psi,\Delta B\Psi)_{\cP}\bigr|.
\end{align}
Taking square of the above and applying the inequality (\ref{eq:CS})
with $\phi=\Delta A\Psi$ and $\psi=\Delta B\Psi$ we obtain
\begin{align}
\bigl|Q_{\Gaminf}(\Psi,[\Delta A,\Delta B]\Psi)_{\cP}\bigr|^{2}&\leq
 4\bigl|Q_{\Gaminf}(\Delta A\Psi,\Delta B\Psi)_{\cP}\bigr|^{2}\notag\\
&\leq 4 Q_{\Gaminf}(\Delta A\Psi,\Delta A\Psi)_{\cP}\,
 Q_{\Gaminf}(\Delta B\Psi,\Delta B\Psi)_{\cP}\notag\\
&=4 Q_{\Gaminf}(\Psi,(\Delta A)^{2}\Psi)_{\cP}\,
 Q_{\Gaminf}(\Psi,(\Delta B)^{2}\Psi)_{\cP}.
\end{align}
By the definition of $\cP$-expectation value (\ref{eq:expv}) and
the commutation relation (\ref{eq:dAdB}), we finally obtain
\begin{align}
\langle(\Delta\hat{A})^{2}\rangle_{\cP}\,\langle(\Delta\hat{B})^{2}
 \rangle_{\cP}\geq\frac{\hbar}{4}\bigl|\langle\hat{C}\rangle_{\cP}
 \bigr|^{2},\qquad\Delta A\Psi,\Delta B\Psi\in\fL_{+}\text{ or }
 \fL_{-},
\label{eq:uncer}
\end{align}
which can be regarded as an uncertainty relation in $\cP\cT$-symmetric
quantum theory.

\section{Classical-Quantum Correspondence}
\label{sec:cqco}

In this section, we shall investigate and discuss classical-quantum
correspondence in the $\cP\cT$-symmetric theory.
The classical equations of motion subjected to
$\cP\cT$-symmetric potentials were analyzed in detail in
Refs.~\cite{BBM99,Na04,BCDM06}. Since $\cP\cT$-symmetric potentials
are generally non-real, the classical motion were considered in
a complex plane. The results indicates a strong correlation between
classical and quantum systems especially from the viewpoint of
$\cP\cT$ symmetry breaking though up to now no correspondence
principle between them in $\cP\cT$-symmetric theory has been
explicitly established. The purposes of this section are first
to establish the $\cP\cT$-symmetric version of the Ehrenfest's
theorem and then to discuss its consequences. As we will see, it
provides a novel correspondence to classical systems
completely different from the conventional one in the literature.

Let us first examine the time derivative of the $\cP$-expectation
value of the position operator~$\hat{x}$:
\begin{align}
\frac{\rmd}{\rmd t}\langle\hat{x}\rangle_{\cP}&=\frac{\rmd}{\rmd t}
 \int_{-\infty}^{\infty}\rmd x\,\Psi^{\ast}(-\zeta^{\ast}(x),t)
 x\Psi(\zeta(x),t)\notag\\
&=\int_{-\infty}^{\infty}\rmd x\left[\frac{\del\Psi^{\ast}
 (-\zeta^{\ast}(x),t)}{\del t}x\Psi(\zeta(x),t)+\Psi^{\ast}
 (\zeta^{\ast}(x),t)x\frac{\del\Psi(\zeta(x),t)}{\del t}\right].
\end{align}
Applying Eqs.~(\ref{eq:tdSch}) and (\ref{eq:tdSch'}), and integrating
by parts, we obtain
\begin{align}
\frac{\rmd}{\rmd t}\langle\hat{x}\rangle_{\cP}&=\frac{\hbar}{2m\rmi}
 \int_{-\infty}^{\infty}\rmd x\left[\frac{\del^{2}\Psi^{\ast}
 (-\zeta^{\ast}(x),t)}{\del x^{2}}x\Psi(\zeta(x),t)-\Psi^{\ast}
 (-\zeta^{\ast}(x),t)x\frac{\del^{2}\Psi(\zeta(x),t)}{\del x^{2}}
 \right]\notag\\
&=-\frac{\hbar}{2m\rmi}\int_{-\infty}^{\infty}\rmd x\,\left[
 \frac{\del\Psi^{\ast}(-\zeta^{\ast}(x),t)}{\del x}\Psi(\zeta(x),t)
 -\Psi^{\ast}(-\zeta^{\ast}(x),t)\frac{\del\Psi(\zeta(x),t)}{\del x}
 \right]\notag\\
&=\frac{\hbar}{m\rmi}\int_{-\infty}^{\infty}\rmd x\,\Psi^{\ast}
 (-\zeta^{\ast}(x),t)\frac{\del\Psi(\zeta(x),t)}{\del x}.
\label{eq:eom1}
\end{align}
Next, the time derivative of the $\cP$-expectation value of
the momentum operator $\hat{p}$ reads
\begin{align}
\frac{\rmd}{\rmd t}\langle\hat{p}\rangle_{\cP}&=-\rmi\hbar
 \frac{\rmd}{\rmd t}\int_{-\infty}^{\infty}\rmd x\,\Psi^{\ast}
 (-\zeta^{\ast}(x),t)\frac{\del\Psi(\zeta(x),t)}{\del x}\notag\\
&=\rmi\hbar\int_{-\infty}^{\infty}\rmd x\,\left[\frac{\del\Psi^{\ast}
 (-\zeta^{\ast}(x),t)}{\del x}\frac{\del\Psi(\zeta(x),t)}{\del t}
 -\frac{\del\Psi^{\ast}(-\zeta^{\ast}(x),t)}{\del t}\frac{\del\Psi
 (\zeta(x),t)}{\del x}\right].
\end{align}
Again, applying Eqs.~(\ref{eq:tdSch}) and (\ref{eq:tdSch'}), and
integrating by parts we obtain
\begin{align}
\frac{\rmd}{\rmd t}\langle\hat{p}\rangle_{\cP}
 &=\int_{-\infty}^{\infty}\rmd x\,\left[\frac{\del\Psi^{\ast}
 (-\zeta^{\ast}(x),t)}{\del x}V(x)\Psi(\zeta(x),t)+\Psi^{\ast}
 (-\zeta^{\ast}(x),t)V(x)\frac{\del\Psi(\zeta(x),t)}{\del x}
 \right]\notag\\
&=-\int_{-\infty}^{\infty}\rmd x\,\Psi^{\ast}(-\zeta^{\ast}(x),t)
 \frac{\rmd V(x)}{\rmd x}\Psi(\zeta(x),t).
\label{eq:eom2}
\end{align}
Hence from Eqs.~(\ref{eq:eom1}) and (\ref{eq:eom2}) we obtain
a set of equations of motion:
\begin{align}
m\frac{\rmd\langle\hat{x}\rangle_{\cP}}{\rmd t}=\langle\hat{p}
 \rangle_{\cP},\qquad \frac{\rmd\langle\hat{p}\rangle_{\cP}}{\rmd t}
 =-\bigl\langle V'(\hat{x})\bigr\rangle_{\cP},
\label{eq:Ehren}
\end{align}
which can be regarded as an alternative to the \emph{Ehrenfest's
theorem} in ordinary Hermitian quantum mechanics. The most
important point is that in the case of $\cP\cT$-symmetric systems
it holds for the expectation values with respect to the indefinite
$\cP$-metric in $L_{\cP}^{2}$ but not with respect to the positive
definite metric in $L^{2}$. As we will show in what follows, it
leads to novel consequences and features of the classical-quantum
correspondence in $\cP\cT$-symmetric theory.

First of all, we recall the fact that both the position and
momentum operators are anti-$\cP$-Hermitian in the Krein space,
(\ref{eq:adjx}) and (\ref{eq:adjp}). We can easily check that
for any $\cP\cT$-symmetric Hamiltonian the first derivative of
the potential $V'(x)$ is also anti-$\cP$-Hermitian:
\begin{align}
Q_{\Gaminf}(\phi, V'(x)^{c}\psi)_{\cP}
&=\int_{-\infty}^{\infty}\rmd x\,\frac{\rmd V^{\ast}(-x)}{\rmd (-x)}
 \phi^{\ast}(-\zeta^{\ast}(x))\psi(\zeta(x))\notag\\
&=-\int_{-\infty}^{\infty}\rmd x\,\phi^{\ast}(-\zeta^{\ast}(x))
 \frac{\rmd V(x)}{\rmd x}\psi(\zeta(x))\notag\\
&=-Q_{\Gaminf}(\phi, V'(x)\psi)_{\cP}.
\end{align}
Thus, all the $\cP$-expectation values appeared in the equations
of motion (\ref{eq:Ehren}), which should correspond to
the classical quantities, are purely imaginary, cf.
Eq.~(\ref{eq:ReIm}).

This consequence naturally let us consider the real quantities
$x_{I}$ and $p_{I}$ as classical `canonical' coordinates defined by
\begin{align}
x(t)=-\rmi x_{I}(t)\in\rmi\bbR,\qquad
 p(t)=-\rmi p_{I}(t)\in\rmi\bbR.
\label{eq:cc1}
\end{align}
Regarding the potential term, we first note that any
$\cP\cT$-symmetric potential satisfying $V^{\ast}(-x)=V(x)$
($x\in\bbR$) can be expressed as
\begin{align}
V(x)=-U(\rmi x),
\end{align}
where $U$ is a real-valued function on $\bbR$, namely,
$U:\bbR\to\bbR$. To show this, we begin with the Laurant expansion
of the potential function:
\begin{align}
V(z)=\sum_{n=-\infty}^{\infty}a_{n}z^{n},\qquad a_{n},z\in\bbC.
\end{align}
Then $\cP\cT$-symmetry of $V$ on $\bbR$ is equivalent to
\begin{align}
a_{n}^{\ast}=(-1)^{n}a_{n},\qquad\forall n\in\bbZ,
\end{align}
that is, $a_{n}$ is real (purely imaginary) for all even (odd) integer
$n$. Thus, without loss of generality we can put for all $n$
\begin{align}
a_{n}=\rmi^{n}b_{n},\qquad b_{n}\in\bbR.
\end{align}
Hence, the $\cP\cT$-symmetric potential reads
\begin{align}
V(x)=\sum_{n=-\infty}^{\infty}b_{n}(\rmi x)^{n}\equiv -U(\rmi x),
\label{eq:relVU}
\end{align}
where $U(y)=-\sum_{n}b_{n}y^{n}$ is in fact a real-valued function
on $\bbR$. As a result, we have in particular
\begin{align}
\frac{\rmd V(x)}{\rmd x}=-\frac{\rmd U(\rmi x)}{\rmd x}
 =-\rmi U'(\rmi x).
\end{align}

Therefore, if we assume the classical-quantum correspondence for
the purely imaginary quantities as
\begin{align}
\langle\hat{x}\rangle_{\cP}\longleftrightarrow x(t)=-\rmi x_{I}(t),
 \qquad
 \langle\hat{p}\rangle_{\cP}\longleftrightarrow p(t)=-\rmi p_{I}(t),
\label{eq:cc2}
\end{align}
the equations of motion for the real `canonical' coordinates
$x_{I}$ and $p_{I}$, which correspond to Eq.~(\ref{eq:Ehren}),
read\footnote{We can choose $x_{I}$ and $p_{I}$ by, e.g., putting
$p(t)=\rmi p_{I}(t)$ instead of the second one in (\ref{eq:cc1})
and (\ref{eq:cc2}) so that their corresponding quantum operators
maintain the canonical commutation relation
$[\hat{x}_{I},\hat{p}_{I}]=\rmi\hbar$. In this case, each of
the equations of motion in (\ref{eq:ceom}) holds with the reversed
sign. But the combined form $m(\rmd^{2}x_{I}/\rmd t^{2})=-U'(x_{I})$
is invariant.}
\begin{align}
m\frac{\rmd x_{I}(t)}{\rmd t}=p_{I}(t),\qquad
 \frac{\rmd p_{I}(t)}{\rmd t}=-U'(x_{I}).
\label{eq:ceom}
\end{align}
That is, they constitute a real dynamical system subject to the
real-valued potential $U$.

This consequence is quite striking. Although $\cP\cT$-symmetric
quantum potentials are complex and supports of square integrable
wave functions are complex contours in general, we can establish
a correspondence of such a $\cP\cT$-symmetric quantum system to
a real classical system. Conversely, for every classical dynamical
system described by a real potential $U(x)$, we can construct
the corresponding $\cP\cT$-symmetric quantum potential $V(x)$
through the relation (\ref{eq:relVU}). In what follows, we exhibit
several $\cP\cT$-symmetric quantum potentials $V$ in the literature
and the corresponding classical potentials $U$ as examples.

\begin{itemize}

\item Example 1 in Ref.~\cite{BBM99}.
 \begin{align}
 V(x)&=x^{2K}(\rmi x)^{\epsilon}.
 \label{eq:ex1}\\
 U(x)&=(-1)^{K+1}x^{2K+\epsilon}.
 \end{align}

\item Example 2 in Ref.~\cite{DDT01a}.
 \begin{align}
 V(x)&=-(\rmi x)^{2M}-\alpha (\rmi x)^{M-1}+\frac{l(l+1)}{x^{2}}.\\
 U(x)&=x^{2M}+\alpha x^{M-1}+\frac{l(l+1)}{x^{2}}.
 \end{align}

\item Example 3 in Ref.~\cite{Zn99c}.
 \begin{align}
 V(x)&=-\omega^{2}\rme^{4\rmi x}-D\rme^{2\rmi x}.\\
 U(x)&=\omega^{2}\rme^{4x}+D\rme^{2x}.
 \end{align}

\item Example 4 in Ref.~\cite{FGRZ99}.
 \begin{align}
 V(x)&=-(\rmi\sinh x)^{\alpha}(\cosh x)^{\beta}.\\
 U(x)&=(\sin x)^{\alpha}(\cos x)^{\beta}.
 \end{align}

\item Example 5 in Ref.~\cite{BDM99}.
 \begin{align}
 V(x)&=\rmi (\sin x)^{2N+1}.\\
 U(x)&=(-1)^{N+1}(\sinh x)^{2N+1}.
 \end{align}

\end{itemize}
For an arbitrary operator $\hat{O}(t)$ which can have explicit
dependence on the time variable $t$, we can also establish
the correspondence. We consider the time derivative of
the $\cP$-expectation value of $\hat{O}$:
\begin{align}
\rmi\hbar\frac{\rmd}{\rmd t}\langle\hat{O}(t)\rangle_{\cP}
=&\,\rmi\hbar\frac{\rmd}{\rmd t}\int_{-\infty}^{\infty}\rmd x\,
 \Psi^{\ast}(-\zeta^{\ast}(x),t)O(t)\Psi(\zeta(x),t)\notag\\
=&\,\rmi\hbar\int_{-\infty}^{\infty}\rmd x\left[\Psi^{\ast}
 (-\zeta^{\ast}(x),t)O(t)\frac{\del\Psi(\zeta(x),t)}{\del t}
 +\frac{\del\Psi^{\ast}(-\zeta^{\ast}(x),t)}{\del t}O(t)
 \Psi(\zeta(x),t)\right]\notag\\
&{}+\rmi\hbar\int_{-\infty}^{\infty}\rmd x\,\Psi^{\ast}(-\zeta^{\ast}
 (x),t)\frac{\del\hat{O}(t)}{\del t}\Psi(\zeta(x),t).
\label{eq:O't}
\end{align}
By virtue of Eqs.~(\ref{eq:tdSch}) and (\ref{eq:tdSch'}), and
the transposition symmetry of the Hamiltonian and the relation
(\ref{eq:trans}), the term in the second line of Eq.~(\ref{eq:O't})
reads
\begin{align}
&\int_{-\infty}^{\infty}\rmd x\,\bigl\{\Psi^{\ast}(-\zeta^{\ast}
 (x),t)O(t)H\Psi(\zeta(x),t)-[H\Psi^{\ast}(-\zeta^{\ast}(x),t)]
 O(t)\Psi(\zeta(x),t)\bigr\}\notag\\
&=\int_{-\infty}^{\infty}\rmd x\,\Psi^{\ast}(-\zeta^{\ast}(x),t)
 (O(t)H-HO(t))\Psi(\zeta(x),t).
\end{align}
Hence, we obtain the generalized Ehrenfest's theorem in
$\cP\cT$-symmetric quantum theory:
\begin{align}
\rmi\hbar\frac{\rmd\langle\hat{O}(t)\rangle_{\cP}}{\rmd t}
 =\bigl\langle [\hat{O}(t),\hat{H}]\bigr\rangle_{\cP}
 +\rmi\hbar\biggl\langle\frac{\del\hat{O}(t)}{\del t}
 \biggr\rangle_{\cP},
\end{align}
where we note again that the formula relates the quantities defined
in terms of the $\cP$-expectation values.

\section{Additional Restrictions on $\cP$-Self-Adjointness}
\label{sec:KH}

Although we have now established the classical-quantum correspondence
in our framework, it is not yet sufficient for the theory to be
physically acceptable. To see this, we shall first review the roles of
self-adjointness in ordinary quantum theory, and then come back to
consider our case. Completeness of eigenvectors is the central issue
in this section. For the later discussions, we introduce the following
notation:
\begin{align}
\fE(A)&=\overline{\braket{\fS_{\lambda}(A)\bigm| \lambda\in
 \sigma_{p}(A)}},\\
\fE_{0}(A)&=\overline{\braket{\Ker(A-\lambda I)\bigm|
 \lambda\in\sigma_{p}(A)}}.
\end{align}
That is, $\fE(A)$ ($\fE_{0}(A)$) is the completion of the vector
space spanned by all the root vectors (eigenvectors) of the operator
$A$, respectively. By definition, $\fE_{0}(A)\subset\fE(A)$.

In ordinary quantum theory, it is crucial that any state vector
in the Hilbert space $L^{2}$ can be expressed as a linear combination
of the eigenstates of the Hamiltonian or physical observables under
consideration. However, this property, called \emph{completeness},
is so frequently employed in vast areas of applications without
any doubt that one may forget the fact that it is guaranteed by
the self-adjointness of the operators. The mathematical theorem
which ensures the completeness of the eigenvectors of
a self-adjoint operator and the existence of an eigenbasis is
the following (Lemma {4.2.7} in \cite{AI89}):
\begin{The}
If $A$ is a self-adjoint operator in a Hilbert space $\fH$ with
a spectrum having no more than a countable set of points of
condensation, then $\fE_{0}(A)=\fH$, and in $\fH$ there is
an orthonormalized basis composed of the eigenvectors of
the operator $A$.
\label{th:self}
\end{The}
We thus emphasize that the postulate of self-adjointness of physical
observable operators in ordinary quantum theory is crucial not
only for the reality of their spectrum but also for the completeness
of their eigenvectors and the existence of a basis composed of them.
Reminded by this important fact, we shall next consider the situation
of $\cP\cT$-symmetric quantum theory defined in the Krein space.

Unfortunately, it has been known that the system of even the root
vectors of a $J$-self-adjoint operator does not generally span
a dense set of the whole Krein space, and more strikingly, that
completeness of the system of the eigenvectors does not guarantee
the existence of a basis composed of such vectors (cf. Section {4.2}
in \cite{AI89}).
Therefore, that the completeness of the eigenvectors (or at worst,
of root vectors) and the existence of a basis composed of them in
the Krein space $L_{\cP}^{2}$ would be inevitable for the theory to
be physically acceptable leads to the following:
\begin{description}
\item[Criterion 3] Every physically acceptable $\cP$-self-adjoint
 operator must admit a complete basis composed of its eigenvectors,
 or at worst, of its root vectors.
\end{description}
Fortunately, we have found that there exists (at least) one, among
subclasses of $J$-self-adjoint operators, which can fulfil Criteria
1--3, namely, the so-called class $\bK(\bH)$~\cite{Az76}. For
the preciseness, we shall present in what follows the mathematical
definitions (cf. Definitions {2.4.2}, {2.4.18}, {3.5.1}, and
{3.5.10} in Ref.~\cite{AI89}). To define the class $\bK(\bH)$,
we first need the following classes of operators:
\begin{Def}
An operator $V$ in a Krein space $\fH_{J}$ with a $J$-metric
$Q(\cdot,\cdot)_{J}$ is said to be \emph{$J$-non-contractive}
if $Q(V\phi,V\phi)_{J}\geq Q(\phi,\phi)_{J}$ for all $\phi\in\fD(V)$.
A continuous $J$-non-contractive operator $V$ with $\fD(V)=\fH_{J}$
is said to be \emph{$J$-bi-non-contractive} if $V^{c}$ is also
$J$-non-contractive.
\end{Def}
\begin{Def}
A bounded operator $T$ is said to \emph{belong to the class}
$\bH$, if it has at least one pair of invariant maximal
non-negative and non-positive subspaces $\fL_{+}\in\fM^{+}$ and
$\fL_{-}\in\fM^{-}$, and if every maximal semi-definite subspace
$\fL_{\pm}\in\fM^{\pm}$ invariant to $T$ belongs to the class
$h^{\pm}$ respectively.
\end{Def}
With these concepts, the class $\bK(\bH)$ is defined as
\begin{Def}
A family of operators $\cA=\{A\}$ is said to \emph{belong to
the class} $\bK(\bH)$ if every operator $A\in\cA$ with
$\rho(A)\cap\bbC^{+}\neq\emptyset$ commutes with
a $J$-bi-non-contractive operator $V_{0}$ of the class $\bH$.
\end{Def}
An important consequence of these definitions can be roughly
described as follows. By virtue of the commutativity with
$V_{0}$, the structure of invariant subspaces of each
$A\in\bK(\bH)$ is mostly  inherited from that of $V_{0}$,
and when $A$ is $J$-self-adjoint also from that of $V_{0}^{c}$.
On the other hand, both of $V_{0}$ and $V_{0}^{c}$ are
$J$-bi-non-contractive and belong to the class $\bH$. In particular,
if $\fL_{+}\in\fM^{+}\cap h^{+}$ ($\fL_{-}\in\fM^{-}\cap h^{-}$)
is an invariant maximal semi-definite subspace of $V_{0}$, then
$\fL_{+}^{[\perp]}\in\fM^{-}\cap h^{-}$
($\fL_{-}^{[\perp]}\in\fM^{+}\cap h^{+}$)
is an invariant maximal semi-definite subspace of $V_{0}^{c}$,
respectively. Hence, a $J$-self-adjoint operator $A\in\bK(\bH)$ can
have an invariant maximal dual pair $(\fL_{+},\fL_{+}^{[\perp]})$ or
$(\fL_{-}^{[\perp]},\fL_{-})$ with $\fL_{\pm},\fL_{\mp}^{[\perp]}
\in\fM^{\pm}\cap h^{\pm}$. Therefore, a $\cP$-self-adjoint
Hamiltonian of the class $\bK(\bH)$ in particular can meet
Criteria 1 and 2 corresponding to Case 1 in Section~\ref{sec:time}.
For a more rigorous understanding, trace related mathematical
theorems in the literature.

Another important consequence is that every neutral invariant
subspace of $A\in\bK(\bH)$ can have at most a \emph{finite}
dimensionality. Then, for every $J$-self-adjoint operator $A$ of
the class $\bK(\bH)$ the Krein space $\fH_{J}$ admits
a $J$-orthogonal decomposition into invariant subspaces of $A$ as
(cf. Section {3.5.6} in \cite{AI89})
\begin{align}
\fH_{J}=\mathop{[\dotplus]}_{i=1}^{\kappa_{1}}\left[
 \fS_{\lambda_{i}}(A)\dotplus\fS_{\lambda_{i}^{\ast}}(A)\right]
 [\dotplus]\fH'_{J},
\label{eq:deco1}
\end{align}
where $\kappa_{1}$ is a \emph{finite} number, and
$\lambda_{i}\not\in\bbR$ are normal non-real eigenvalues of $A$.
Relative to the above decomposition of the space, the operator $A$
has block diagonal form:
\begin{align}
A=\left(
 \begin{array}{rrrr}
 A_{1} &        &                &  \\
       & \ddots &                &  \\
       &        & A_{\kappa_{1}} &  \\
       &        &                & A'
 \end{array}\right),
\end{align}
where $A_{i}=A|_{\fS_{\lambda_{i}}\dotplus\fS_{\lambda_{i}^{\ast}}}$
and $A'=A|_{\fH'_{J}}$. The spectrum of the operator
$A'$ is real, $\sigma(A')\subset\bbR$, and there is at most a
\emph{finite} number $k$ of real eigenvalues $\mu_{i}$ for
which the eigenspaces $\Ker (A^{(\prime)}-\mu_{i}I)$ are
degenerate. The set of such points $\{\mu_{i}\}_{1}^{k}$ is called
the \emph{set of critical points} and denoted by $s(A)$.
In particular, the number $\kappa_{2}$ of non-semi-simple
real eigenvalues is also finite with $\kappa_{2}\leq k$
(cf. Proposition~\ref{th:nsemi}).

It is evident that when $\kappa_{1}=0$, the operator $A$ has no
non-real eigenvalues and thus $\cP\cT$ symmetry is unbroken
(in the weak sense).
However, as we have discussed in Section~\ref{sec:spec}
the existence of neutral eigenvectors in the real sector,
and thus the value of $k$, has no direct relation to
the ill-definiteness and breakdown of $\cP\cT$ symmetry.
In particular, we should note that $\kappa_{1}=k=0$ does not
guarantee unbroken $\cP\cT$ symmetry in the strong sense;
for a degenerate real semi-simple eigenvalue $\lambda$
($m_{\lambda}^{(a)}=m_{\lambda}^{(g)}>1$) the corresponding
eigenspace can be non-degenerate (hence it does not contribute
to the values of either $\kappa_{1}$ and $k$) but $\cP\cT$
symmetry can be ill-defined (cf. Table~\ref{tb:asp}).
Hence, the class $\bK(\bH)$ cannot characterize unbroken $\cP\cT$
symmetry perfectly, but it can certainly exclude a pathological
case where an infinite number of neutral eigenvectors emerges.

Let us now come back to the central problems in this section.
Regarding the completeness and existence of a basis, the following
theorem has been proved:\footnote{Here we omit the assertion on
the existence of a $p$-basis in the original for the simplicity.}
\begin{The}[Azizov \cite{AI89,Az80}]
Let $A$ be a continuous $J$-self-adjoint operator of the class
$\bK(\bH)$ in a Krein space $\fH_{J}$, and let $\sigma(A)$ have
no more than a countable set of points of condensation. Then:
 \begin{enumerate}

 \item $\dim\fH_{J}/\fE(A)\leq\dim\fH_{J}/\fE_{0}(A)<\infty$;

 \item $\fE_{0}(A)=\fH_{J}$ if and only if $s(A)=\emptyset$
  and $\fS_{\lambda}(A)=\Ker(A-\lambda I)$ when
  $\lambda\neq\lambda^{\ast}$;

 \item $\fE(A)=\fH_{J}$ if and only if $\braket{\fS_{\lambda}(A)
  \bigm| \lambda\in s(A)}$ is a non-degenerate subspace:

 \item if $\fE_{0}(A)=\fH_{J}$ (respectively, $\fE(A)=\fH_{J}$),
  then there is in $\fH_{J}$ an almost $J$-ortho\-normalized (Riesz)
  basis composed of eigenvectors (respectively, root vectors) of
  the operator $A$;

 \item if $\fE_{0}(A)=\fH_{J}$, then there is in $\fH_{J}$ a
  $J$-orthonormalized (Riesz) basis composed of eigenvectors of
  the operator $A$ if and only if $\sigma(A)\subset\bbR$.

 \end{enumerate}
\label{th:azi}
\end{The}
Comparing Theorems~\ref{th:self} and \ref{th:azi}, one easily
recognizes the complicated situation in the case of Krein spaces,
and, in particular, the fact that even the restriction to the class
$\bK(\bH)$ does not automatically guarantee the completeness of
the system of eigenvectors or root vectors.

Let us first discuss consequences of the theorem for the case
$\fE(A)=\fH_{J}$. By virtue of Corollary~\ref{th:nond} and the third
assertion in Theorem~\ref{th:azi}, one of the sufficient conditions
for $\fE(A)=\fH_{J}$ is that all the real eigenvalues $\mu_{i}$
belonging to $s(A)$ are normal. In this case, they must not be
semi-simple; otherwise, $\fS_{\mu_{i}}(A)=\Ker(A-\mu_{i}I)$ is
non-degenerate, which contradicts $\mu_{i}\in s(A)$. Hence, we have
$k=\kappa_{2}$ and can further decompose the space (\ref{eq:deco1}) as
\begin{align}
\fH_{J}=\mathop{[\dotplus]}_{i=1}^{\kappa_{1}}\left[
 \fS_{\lambda_{i}}(A)\dotplus\fS_{\lambda_{i}^{\ast}}(A)\right]
 \mathop{[\dotplus]}_{i=1}^{\kappa_{2}}\fS_{\mu_{i}}(A)
 [\dotplus]\fH''_{J},
\end{align}
where $\mu_{i}\in s(A)$ and thus each $\fS_{\mu_{i}}$ contains
at least one neutral eigenvector, and the spectrum of
$A''\equiv A|_{\fH''_{J}}$ is real $\sigma(A'')\subset\bbR$,
and in particular, all the eigenvalues of $A''$ are real and
semi-simple with non-degenerate eigenspaces.
Then the fourth assertion guarantees that the system of the root
vectors can always constitute an almost $J$-orthonormalized basis,
that is, it is the union of a finite subset of vectors
$\{f_{i}\}_{1}^{n}$ and a $J$-orthonormalized subset
$\{e_{i}\}_{1}^{\infty}$ satisfying $Q(e_{i},e_{j})_{J}=\delta_{ij}$
or $-\delta_{ij}$, these two subsets being $J$-orthogonal to
one another (Definition {4.2.10} in \cite{AI89}) such that
\begin{align}
\fH_{J}=\overline{\langle f_{1},\dots,f_{n}\rangle [\dotplus]
 \langle e_{1},e_{2},\ldots\rangle}.
\label{eq:alstJ}
\end{align}

Next, we shall consider the most desirable case $\fE_{0}(A)=\fH_{J}$
where eigenvectors of $A$ span a dense subset of the whole space
$\fH_{J}$. The second assertion in Theorem~\ref{th:azi} means that
it is the case if and only if all the eigenvalues of $A$ are
semi-simple with no degenerate eigenspaces. But a degenerate
eigenspace belonging to a real semi-simple eigenvalue can exist
only when the eigenvalue is not normal; from Corollary~\ref{th:nond}
for every normal real semi-simple eigenvalue $\lambda$ the
corresponding eigenspace $\Ker(A-\lambda I)=\fS_{\lambda}(A)$ is
always non-degenerate. Hence, so long as all the real eigenvalues
are normal in this case, we always have
\begin{align}
\fH_{J}=\fE_{0}(A)=\mathop{[\dotplus]}_{i=1}^{\kappa_{1}}
 \overline{
 \left[\Ker(A-\lambda_{i}I)\dotplus\Ker(A-\lambda_{i}^{\ast}I)
 \right]\mathop{[\dotplus]}_{\lambda\subset\bbR}\Ker(A-\lambda I)},
\label{eq:alln2}
\end{align}
which can be regarded as a special case of Eq.~(\ref{eq:alln})
in Proposition~\ref{th:alln}. Here we recall the fact that all
the eigenvectors corresponding to non-real eigenvalues are neutral
and thus cannot be elements of a $J$-orthonormalized basis
$\{e_{i}\}$ satisfying $Q(e_{i},e_{j})_{J}=\delta_{ij}$ or
$-\delta_{ij}$. Thus, in the case of Eq.~(\ref{eq:alln2}),
the number $n$ in Eq.~(\ref{eq:alstJ}) is given by
\begin{align}
n=\sum_{i=1}^{\kappa_{1}}\left[m_{\lambda_{i}}^{(a)}(A)
 +m_{\lambda_{i}^{\ast}}^{(a)}(A)\right]=2\sum_{i=1}^{\kappa_{1}}
 m_{\lambda_{i}}^{(g)}(A).
\end{align}
Hence, if furthermore all the eigenvalues are real, namely,
$\kappa_{1}=0$, we have $n=0$ and thus the system of eigenvectors
can form a $J$-orthonormalized basis, as is indeed ensured by
the fifth assertion.

We now understand that for an arbitrary continuous $\cP$-self-adjoint
operator $A$ of the class $\bK(\bH)$ the completeness
$\fE_{0}(A)=L_{\cP}^{2}$ and the existence of a basis composed of
the eigenvectors of $A$ in $L_{\cP}^{2}$ are guaranteed if and only
if all the eigenvalues are semi-simple and there is no degenerate
eigenspaces in the real sector, irrespective of the existence
of non-real eigenvalues, or in other words, irrespective of whether
$\cP\cT$ symmetry is spontaneously broken. But the latter condition
of the non-degeneracy is always guaranteed unless there appears a
non-normal real eigenvalue, as we have just discussed. Thus, in most
of the cases we would not need to resort to the quotient-space
prescription proposed in the previous paper~\cite{Ta06b}. Therefore,
a remaining problem is how to deal with neutral eigenvectors
belonging to non-real eigenvalues when $\cP\cT$ symmetry is
spontaneously broken. In our previous paper \cite{Ta06b}, we have
proposed the possibility to interpret them as physical states
describing unstable decaying states (and their `spacetime-reversal'
states). Until now we have not found any active reason to discard it.
However, the fact that we must always restrict ourselves to
a semi-definite subspace for the probability interpretation would
make the role of these neutral vectors quite restrictive (cf.
Section~\ref{sec:time}).

\section{Discussion and Summary}
\label{sec:disc}

In this work, we have revealed the various general aspects
of $\cP\cT$-symmetric quantum theory defined in the Krein space
$L_{\cP}^{2}$, previously proposed by us in Ref.~\cite{Ta06b}.
The fact that $\cP$-self-adjoint Hamiltonians `favor' the Krein
space $L_{\cP}^{2}$ rather than the Hilbert space $L^{2}$
inevitably led us to formulate a quantum theory in a space with
an indefinite metric. Attempts to quantize a physical system in
an indefinite metric space are traced back to the Dirac's work in
1942 \cite{Di42}. Since then, there have appeared numerous attempts
of this kind in various context (see references cited in
Ref.~\cite{Kl04}, and Ref.~\cite{KO79} for non-Abelian gauge theories).
The significant feature in our case is the conservation law
(\ref{eq:cons}), which ensures that the character (positivity etc.)
of every state vectors remains unchanged in the time evolution.
This, together with the fact that the probability interpretation
is possible only in a semi-definite space, naturally led us to
Criteria 1 and 2. Here we note that they would be valid not only
in our present case but also in any case one would like to quantize
a system in a Krein space. Criterion 3 would be indispensable in
any kind of quantum theory. We have found that there exists a class
of $J$-self-adjoint operators, called the class $\bK(\bH)$, which
can satisfy those 3 criteria.

We note that our quantization scheme of $\cP\cT$-symmetric theory
turns to be completely different from the existing approaches
such as the use of $\cC$ or positive metric operators. The origin
of the difference is twofold. The first reason comes from the
different settings of eigenvalue problems, Eqs.~(\ref{eq:tdSch})
and (\ref{eq:conv1}). Our setting may seem to be strange
especially to those who are familiar with the conventional one.
But it is the conventional setting that makes the physical
interpretation of, e.g., the complex position and momentum quite
difficult when a system cannot be defined on the real line.
It is evident that this difficulty cannot be overcome even if
we express a Hamiltonian in terms of a real variable as
Eq.~(\ref{eq:conv2}); the `Hamiltonian' which determines the
time evolution of state vectors is no longer a Schr\"odinger
operator and it is almost impossible to establish any
correspondence with Newtonian classical dynamical systems
and any reasonable interpretation of physical observables even
for the most fundamental ones such as position and momentum.
The second reason comes from the different choices of linear
spaces where one would like to make a probability interpretation.
In the conventional approaches, we must transform a given
Hamiltonian to another operator acting in a positive definite
Hilbert space. But the transformed operator is in general not
a Schr\"odinger operator and thus they suffer from the same
problem as the one just mentioned above (besides the problem
of unboundedness of metric operators), cf.
Refs.~\cite{MB04,Mo05b,Jo05}. This difficulty is in fact
the central origin of the disputes found in e.g.
Refs.~\cite{BCM06,Mo06a}.

In this respect, we would like to emphasize first that reality of
spectrum, existence of a positive-definite norm, and so on, are
neither sufficient nor necessary conditions for a quantum-like
theory to be physically acceptable. Crucial viewpoints must be
put on the possibility whether we can assign a reasonable physical
interpretation for each consequence of the theory and on
the consistency of the interpretation with experimental results.
Regarding the former point of view, we have not detected so far
any difficulty in physical interpretations, apparent breakdown
or fatal inconsistency in our framework, and
the investigations presented in this paper indicates that
it can stand as another consistent quantum theory.
Therefore, among the different mathematical settings of
the eigenvalue problems for $\cP\cT$-symmetric Hamiltonians
in the literature, our formulation has been shown to possess
the most desirable property as a physical quantum theory,
though the others are still quite interesting as purely
mathematical problems. But we should regard the present
results as a necessary minimum and explore further extensive
studies to see whether the theory would be certainly
free from any insurmountable discrepancy.

The classical-quantum correspondence in $\cP\cT$-symmetric theory
we have established in Section~\ref{sec:cqco} suggests that
we should regard $\cP\cT$-symmetric quantum theory as another
quantization scheme rather than a generalization of traditional
Hermitian quantum theory. That is, for a given real classical
potential $U(x)$ we associate the real Hermitian operator
$U(\hat{x})$ acting in the Hilbert space $L^{2}$ in
the traditional scheme, while we associate the complex
$\cP$-Hermitian operator $V(\hat{x})$, obtained through
the relation (\ref{eq:relVU}), acting in the Krein space
$L_{\cP}^{2}$ in the $\cP\cT$-symmetric scheme. But certainly
some of classical potentials $U(x)$ would only admit
a $\cP\cT$-symmetric quantization but not a Hermitian one due to
the lack of normalizable eigenfunctions for $U(x)$ on $x\in\bbR$.
An intriguing situation can arise when a given classical potential
admits both quantization schemes. As a physical theory, which of
them we should take must be of course determined by the comparison
between theoretical prediction and experiment. From this point of
view, it is quite interesting to examine the cases where the two
different quantum models constructed from a single classical
system predict different physical consequences.

The classical systems obtained from the classical-quantum
correspondence are completely real. On the other hand,
the $\cP\cT$-symmetric complex classical systems investigated
in Refs.~\cite{BBM99,Na04,BCDM06} have (at least until now) no
correspondence principle which relates them to
the $\cP\cT$-symmetric quantum systems. Nevertheless,
the results in the latter references strongly indicate
an intimate relation between complex classical and quantum systems
especially in view of spontaneous $\cP\cT$ symmetry breaking.
We can easily expect that the corresponding real classical systems
in our framework would be insensitive to $\cP\cT$ symmetry breaking
at the quantum level since $\cP$-expectation value of $\cP$-Hermitian
Hamiltonians are always real, and in particular zero for every
eigenstates belonging to non-real energy eigenvalues. Therefore,
it is still quite important to reveal and understand underlying
dynamical relations between $\cP\cT$-symmetric complex classical
and quantum systems.

We have shown in Section~\ref{ssec:pHerm} that the concept of
transposition symmetry plays a key role in connecting $\cP\cT$
symmetry with $\cP$-Hermiticity. On the other hand, as was pointed
out in Ref.~\cite{Si70}, this symmetry underlies the intimate
relation between level crossing phenomena and Jordan anomalous
behavior. The fact that these two significant aspects rely on
the same property would not be accidental. In fact, several
$\cP\cT$-symmetric Hamiltonians characterized by a set of
parameters have non-trivial phase diagrams \cite{BB98a,BBM99,DDT01b},
that is, they have both the symmetric and broken phases of
$\cP\cT$ symmetry in the parameter-space and at the boundary
a level crossing takes place;
a pair of different real eigenvalues in the symmetric phase
degenerates at the boundary and then splits into a complex-conjugate
pair in the broken phase.
Hence, the underlying mechanism of the emergence of a non-trivial
phase diagram of $\cP\cT$ symmetry and that of the Bender--Wu
singularities \cite{BW68,BW69,Si70} would be essentially the same.
In this respect, it is also interesting to note that, the structure
of the energy levels of the potential (\ref{eq:ex1}) shown in
Ref.~\cite{BBM99} indicates that $\epsilon=0$ would be
an accumulation point of the spectral singularity in
the $\epsilon$-plane, which also resembles the structure of
the Bender--Wu singularities despite the totally different roles
of the parameters between them. We further recall the analogous
situation in the Lee model. It was already shown by Heisenberg
in 1957 that at the critical point where the `dipole-ghost' state
emerges, the system exhibits a level crossing and admits an
associated state vector and a zero-norm eigenstate \cite{He57}.

The restriction $\Delta A\Psi,\Delta B\Psi\in\fL_{+}$ or $\fL_{-}$
for the uncertainty relation in Eq.~(\ref{eq:uncer}) suggests that
every operator $O$ corresponding to a physical observable should also
preserve the invariant maximal semi-definite subspaces $\fL_{\pm}$
relative to the Hamiltonian $H$. It would be possible if each operator
$O$ commutes with the operator $V_{0}\in\bH$ which characterizes
the class $\bK(\bH)$ to which the Hamiltonian $H$ belongs. From this
observation, we reach the following postulate for an operator $O$ to
be a physical observable:
\begin{description}
\item[Postulate] A set of physical observables $\cO=\{O\}$ for
 a given Hamiltonian $H\in\bK(\bH)$ is a family of $\cP$-self-adjoint
 or anti-$\cP$-self-adjoint operators belonging to the same class
 $\bK(\bH)$ characterized by the same $V_{0}$, namely,
 $H\cup\cO\in\bK(\bH,V_{0})$.
\end{description}
Then, a natural question is whether we can make a sensible physical
interpretation of the operator $V_{0}$. If it turns out that it is
indeed possible, it might provide a physical reason why we should
restrict ourselves to $\cP$-self-adjoint operators of the class
$\bK(\bH)$. Mathematically, the class $\bK(\bH)$ would not be
a necessary condition for satisfying Criteria 1 and 2. Thus, it
is possible that another class of $J$-self-adjoint operators
which is more suitable for a physical application could be found
in the future.

Regarding the Azizov's theorem in Section~\ref{sec:KH}, we note
the fact that rigorously speaking it applies only to continuous
operators. On the other hand, physical Hamiltonians we are
interested in are usually unbounded. To the best of our knowledge,
an extension of the theorem to unbounded operators has not been
established. We expect most of the assertions would remain valid
also for unbounded operators under a relatively small number of
additional assumptions. We hope this paper would interest and
motivate some mathematicians to study the issue.

\begin{acknowledgments}
 We would like to thank C.~M.~Bender for the valuable discussion on
 the anharmonic oscillator, and A.~van~Tonder for the bibliographic
 information.
 This work was partially supported by the National Science Council
 of the Republic of China under the grant No. NSC-93-2112-M-032-009
 and by the National Cheng-Kung University under the grant
 No. OUA:95-3-2-071.
\end{acknowledgments}


\bibliography{refsels}
\bibliographystyle{npb}



\end{document}